\documentclass[
 reprint,
superscriptaddress,
 amsmath,amssymb,
 aps,
 onecolumn
]{revtex4-2}

\usepackage{graphicx}
\usepackage{dcolumn}
\usepackage{bm}

\usepackage{float}
\usepackage{multirow}
\usepackage{xcolor}
\begin{document}

\preprint{APS/123-QED}

\title{Scalar fields around a loop quantum gravity black hole in de Sitter spacetime: Quasinormal modes, late-time tails and strong cosmic censorship}
\author{Cai-Ying Shao}\email[E-mail: ]{cyshao@hust.edu.cn}\affiliation{MOE Key Laboratory of Fundamental Physical Quantities Measurement, Hubei Key Laboratory of Gravitation and Quantum Physics, PGMF, and School of Physics, Huazhong University of Science and Technology, Wuhan 430074, Hubei, China}
\author{Cong Zhang}\email[E-mail: ]{zhang.cong@mail.bnu.edu.cn}\affiliation{Department Physik, Institut f\"ur Quantengravitation, Theoretische Physik III, Friedrich-Alexander-Universit\"at Erlangen-Nürnberg, Staudtstra{\ss}e 7/B2, 91058 Erlangen, Germany}
\author{Wei Zhang}\email[E-mail: ]{w.zhang@mail.bnu.edu.cn}\affiliation{Key Laboratory of Multiscale Spin Physics, Ministry of Education, and Department of Physics, Beijing Normal University, Beijing 100875, China}
\author{Cheng-Gang Shao}\email[E-mail: ]{cgshao@hust.edu.cn}\affiliation{MOE Key Laboratory of Fundamental Physical Quantities Measurement, Hubei Key Laboratory of Gravitation and Quantum Physics, PGMF, and School of Physics, Huazhong University of Science and Technology, Wuhan 430074, Hubei, China}

\date{\today}

\begin{abstract}
Loop quantum gravity, as one branch of quantum gravity, holds the potential to explore the fundamental nature of black holes.
Recently, according to the quantum Oppenheimer-Snyder model in loop quantum cosmology, a novel loop quantum corrected black hole in de Sitter spacetime has been discovered.
Here, we first investigate the corresponding quasinormal modes and late-time behavior of massless neutral scalar field perturbations based on such a quantum-modified black hole in de Sitter spacetime.
The frequency and time domain analysis of the lowest-lying quasinormal modes is derived by Prony method, Matrix method as well as WKB approximation.
The influences of loop quantum correction, the black hole mass ratio, and the cosmological constant on the quasinormal frequencies are studied in detail.
The late-time behaviors of quantum-modified black holes possess an exponential decay, which is mainly determined not only by the multipole number but also by the cosmological constant.
The impact of loop quantum correction on the late-time tail is negligible, but it has a significant impact on damping oscillation. 
To explore spacetime singularities, we examine the validity of strong cosmic censorship for a near-extremal quantum-modified black hole in de Sitter spacetime. 
As a result, it is found that the strong cosmic censorship is destroyed as the black hole approaches the near-extremal limit, but the violation becomes weaker as the cosmological constant and the loop quantum correction increase.
\end{abstract}

\maketitle

\section{\label{section1}Introduction}
Spacetime singularities, characterized by infinite curvature or density, have been a subject of great interest and curiosity in the fields of gravitation theory and relativistic astrophysics.  According to the singularity theorems proved by Hawking and Penrose, the existence of singularities is unavoidable in generic gravitational collapses. The presence of singularities poses profound challenges to our understanding of the universe within the context of classical general relativity. One specific concern is the existence of naked singularities, which are singularities that are not hidden within a black hole event horizon and thus could be observed by outside observers, breaking down the predictive power of classical general relativity. 

A fine gravitational theory is essentially expected to deal with the problem of spacetime singularities.
From a theoretical standpoint, it is necessary to formulate a theory of quantum gravity (QG) that combines the principles of quantum mechanics and general relativity.
Among the various approaches to QG, loop quantum gravity (LQG) has shown great promise with significant advancements made (see, e.g., \cite{Rovelli:1994ge,Ashtekar:1997fb,han2007fundamental,ashtekar2006quantumnature,ashtekar2011loop,Zhang:2021qul,Zhang:2022vsl} and the references therein). 
By applying the procedure of loop quantization to spherically symmetric black holes, one has gained many insights into the quantum nature of black holes \cite{Chiou:2008nm,PhysRevLett.101.161301,Haggard:2014rza,Christodoulou:2016vny,Ashtekar:2018lag,Zhang:2020qxw,Zhang:2021wex,Lewandowski:2022zce,Husain:2022gwp,Han:2022rsx,Han:2023wxg}, where the singularity of the Schwarzschild black hole is believed to be resolved through the effects of LQG as it should be the case, although the specific detail of how this resolution occurs is scheme dependent.
More interestingly, the quantum geometry effect can make the big bang singularity replaced by a non-singular big bounce~\cite{Stachowiak:2006uh,Ashtekar:2005qt} and provide more effective black hole models~\cite{Modesto:2005zm,Bojowald:2016itl,Chiou:2008eg}.
In particular, with the quantum Oppenheimer-Snyder model in loop quantum cosmology, a new quantum black hole model has been derived most recently \cite{Lewandowski:2022zce}, where the Schwarzschild singularity is resolved by a transition region that contains an inner horizon. 
Given this quantum black hole solution, the quantum geometry modification has been explored in shadows and stability of the black hole~\cite{Zhang:2023okw,Yang:2022btw}, where a sequence of discrepancies was noticed in asymptotically flat spacetimes.
Recent astronomical observations show that there is a mysterious force that urges the universe to expand acceleratedly, which implies the possibility of the existence of a cosmological constant.
Thus, it is necessary to consider the effect of a cosmological constant in the response of quantum-corrected black holes to minor perturbations.

When black holes are slightly disturbed, the resulting evolution is divided into stages: the irregular initial burst determined by the initial conditions, the damped oscillations described by quasinormal modes, and a power-law or exponential tails at late time.
Among them, the quasinormal modes and the late-time tail, which are applied to test the stability of black hole spacetime against small perturbations, carry key information about the related nonperturbed black holes.
Thus, the quasinormal modes have been explored intensively in the test of the no-hair theorem~\cite{Isi:2019aib,Gossan:2011ha,Ota:2019bzl} and validation of modified theories of gravity~\cite{Carson:2020ter,Shao:2023yjx,Bao:2019kgt}.
In addition, the quasinormal modes are crucial in the ringdown waveform of the coalescence of binary systems, which can provide observational evidence of the black holes.
In the wake of developments in science and technology, more broader frequency band of the ringdown phase is speculated to be detected with future space-based observations, which can provide an unparalleled opportunity to test the nature of gravity~\cite{Berti:2015itd,RevModPhys.83.793}.
It would be intriguing and significant to constrain quantum correction parameters of the black holes in LQG with the observed ringdown signals.
On the other hand, the late-time tails can reflect some essential properties of black holes, such as the no-hair theorem and the instability of Cauchy horizons~\cite{prl_scc_qnm,Ori:1999phc,Poisson:1990eh}, which is conducive to understanding the internal structure of black holes.
To this end, we perform a study on quasinormal modes and late-time tails of scalar perturbations for such a loop quantum gravity black hole in de Sitter spacetime.

On the other hand, as the stability of the Cauchy horizon is closely related to the decay rate of dominant quasinormal modes, an important question is whether the strong cosmic censorship conjecture (SCC) is valid in a current physical system.
The SCC proposes that the timelike singularities are not allowed, or can be formulated equivalently as a more rigorous mathematical statement that the Cauchy horizon inside of the black hole is unstable for the generic perturbations and thus inextendible.
Actually, in asymptotically flat spacetimes, the SCC is always valid except for the accelerating black holes~\cite{Destounis:2020pjk,Destounis:2020yav,Zhang10}. However, the validity of SCC will become more complicated in the asymptotically de Sitter spacetimes. A positive cosmological constant leads to an exponential decay of the external perturbations, which can compete with the aforementioned blueshift effect along the Cauchy horizon~\cite{prd_sds_tail,prd_ds_field}.
Thus the validity of the SCC depends on which one will win in the competition.
To be more specific, the SCC has recently been found violated in the nearly extremal charged Reissner-Nordstrom de Sitter (RNdS) black hole by the scalar field~\cite{prl_scc_qnm,prd-scc-charg-scal,prd-scc-charg-Cardoso,cqg-scc-charge-Harvey,prd_scc_higher_dimension}, the fermionic field~\cite{jhep_scc_dirac_highter_dimension,jhep_scc_dirac_rnds,plb-charged-fermions}, and the gravito-electromagnetic field~\cite{jhep_scc_smooth}.
In addition, as to the rotating Kerr de Sitter black hole, the SCC can be respected by the bosonic field perturbations~\cite{prd-scc-de-sitt,jhep-scc-highspacetime}, but violated by the fermionic field perturbation~\cite{epjc-scc-dirac}.
While for the Kerr-Newman de Sitter black hole, the SCC is still violated by both the scalar and fermionic fields~\cite{Casals:2020uxa}.
In this paper, the global structure of such a quantum black hole model resembles that of the charged Reissner-Nordstrom black hole.  
In this sense, the SCC is still plagued potentially by the emergence of the inner Cauchy horizon if one immerses this quantum-modified black hole in de Sitter space.
Thus, we further examine whether the SCC holds for such a quantum-modified black hole in de Sitter spacetime.
In particular, Hollands considered a quantum scalar field in RNdS spacetime and discovered that the quantum effect of the perturbation field can rescue the SCC in semiclassical analysis~\cite{Hollands:2019whz}.
On the contrary, we utilize classical scalar fields in quantum-corrected spacetimes to check the SCC.
Comparing these two semiclassical analyses, we can understand in which case quantum effects have a more significant impact on the SCC.
The remainder of our paper is organized as follows. 
In the next section, we present the quantum Oppenheimer-Snyder model and the corresponding modified metric of the loop quantum black hole in de Sitter spacetime. 
Then we present the equation of motion for a neutral massless scalar field.
We introduce Prony method, Matrix method, and WKB approximation to calculate the quasinormal modes and present the corresponding numerical results in Sec.~\ref{section3}.
In Sec.~\ref{section4}, we illustrate the dynamics of a neutral massless scalar perturbation and explore late-time behaviors of the scalar field for such a quantum-modified black hole.
With the above preparation, the validity of the SCC is checked in Sec.~\ref{section5}.
Finally, the concluding remarks are presented in the last section.

\section{\label{section2}The loop quantum gravity corrected geometry of the black hole in de Sitter space}
Let us follow the precedure introduced in \cite{Lewandowski:2022zce} to get the quantum modified spacetime by considering the quantum Oppenheimer-Snyder model. 
In this model, the entire spacetime is divided into two regions. 
One region comprises a pressureless dust ball with a constant density, and the other region is a vacuum outside the dust ball. 

In the region with dust, we introduce a coordinate $(\tau,\tilde r,\theta,\phi)$ with $0<\tilde r<\tilde r_0$ which adapts the symmetry of the dust ball. 
Then, the metric of the ball takes the form 
\begin{equation}
d s^2_{\rm in}=-d\tau^2+a(\tau)^2 (d\tilde r^2+d\Omega^2),
\end{equation}
where $d\Omega^2=d\theta^2+\sin^2\theta d\phi^2$. 
The dynamics of the scale factor $a(\tau)$ is governed by the LQC modified Friedmann equation
\begin{equation}\label{eq:fridemann}
H^2=\left(\frac{\dot a}{a}\right)^2=\frac{8\pi G}{3}\rho_{\rm tot}(1-\frac{\rho_{\rm tot}}{\rho_c}),\quad \rho_{\rm tot}=\rho_{\rm matter}+\rho_{\Lambda},
\end{equation}
with
\begin{equation*}
\rho_{\rm matter}=\frac{M}{\frac{4}{3}\pi\tilde r_0^3 a^3},\quad \rho_{\Lambda}=\frac{\Lambda}{8\pi G},
\end{equation*}
where the deformation parameter $\rho_c$ denotes the critical density defined as $\rho_c=\sqrt{3}/(32\pi^2\gamma^3 G^2\hbar)$ with the Barbero-Immirzi parameter $\gamma$.
$M$ is the mass of the ball with radius $a(\tau)\tilde r_0$. 
It should be noted that the current work adds a cosmological constant term to the modified Friedmann equation, different from the initial model considered in \cite{Lewandowski:2022zce}.
Eq.(\ref{eq:fridemann}) reverts to the usual Friedmann equation in the classical regime where $\rho\ll \rho_c$. 
However, in the quantum regime where $\rho$ is comparable with $\rho_c$ so that the spacetime curvature becomes Planckian, the deformation term will prevent the matter density $\rho(\tau)$ from reaching infinity which thus prevents the formation of the singularity. 
Indeed, according to Eq.(\ref{eq:fridemann}), at the moment $\tau_b$ with $\rho_{\rm matter}(\tau_b)=\rho_c-\rho_\Lambda$, one has $H=0$, which signifies a change of the dynamics of the ball from the collapsing phase to the expanding phase at $\tau_b$.

In the outside region of the dust ball, we assume the spacetime to be spherically symmetric and static, as done in \cite{Lewandowski:2022zce}. 
We can use the coordinates $(t, r, \theta, \phi)$ to describe this region, which is adapted to the symmetry of the spacetime. 
In this coordinate, the metric of the outside region reads
\begin{equation}
ds^2_{\rm out}=-f(r) dt^2+ g(r)^{-1} d r^2+r^2 d\Omega^2,
\end{equation}
where $f(r)$ and $g(r)$ are two unknown functions to be determined. 
In order to determine the unknown functions $f(r)$ and $g(r)$, we need to find the inner most boundary of the outside region which is glued with the dust ball surface. 
The junction condition for the gluing requires that the reduced 3-metrics and the extrinsic curvatures along the gluing surfaces obtained from the 4-metrics $d s^2_{\rm in}$ and $d s^2_{\rm out}$ respectively are continuous. 
It should be noted that the worldlines $\tau\mapsto (\tau,\tilde r_0,\theta,\phi)$ of each particle on the surface of the dust ball is a timelike geodesic without rotation. 
This implies that the inner most surface of the outside region is also composed of the congruence of freely falling timelike geodesics associated to the metric $ds^2_{\rm out}$.  
Moreover, let $\tau\to (t(\tau),r(\tau),\theta,\phi)$ be a geodesic in the innermost surface of the outside region, with $\tau$ being the length of the geodesic. 
Then, the surfaces are glued by the identification $(\tau,\tilde r_0,\theta,\phi)\sim (t(\tau),r(\tau),\theta,\phi)$. 
The calculation could be simplified by such a junction condition. 

\begin{figure}[H]
\centering
\includegraphics[scale=0.5]{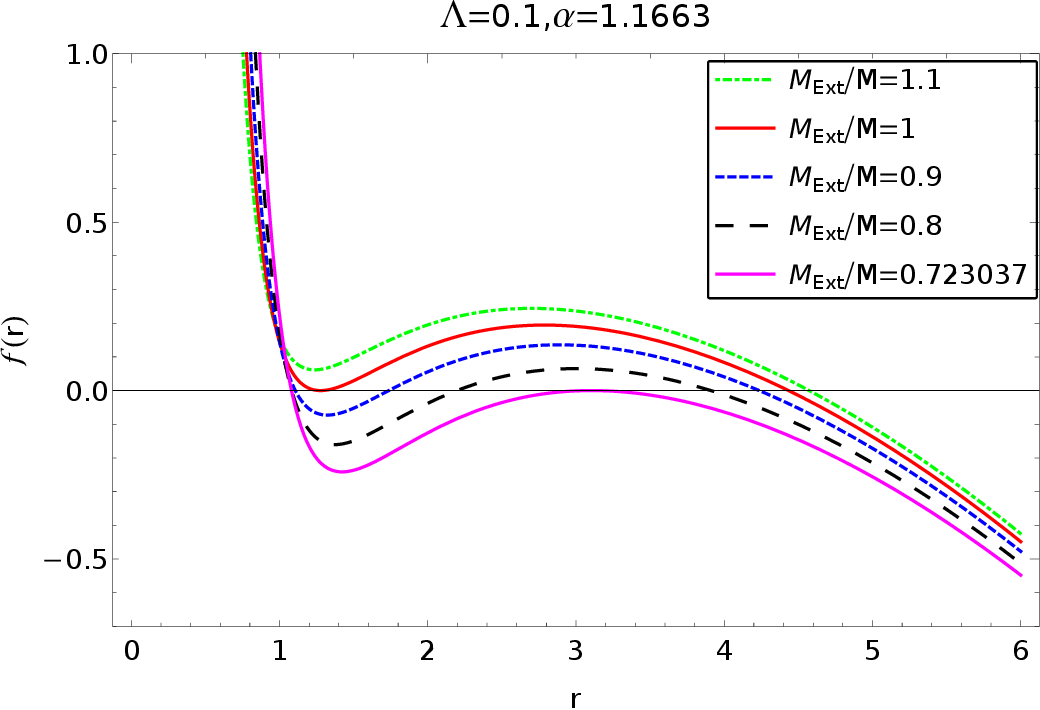}
\includegraphics[scale=0.5]{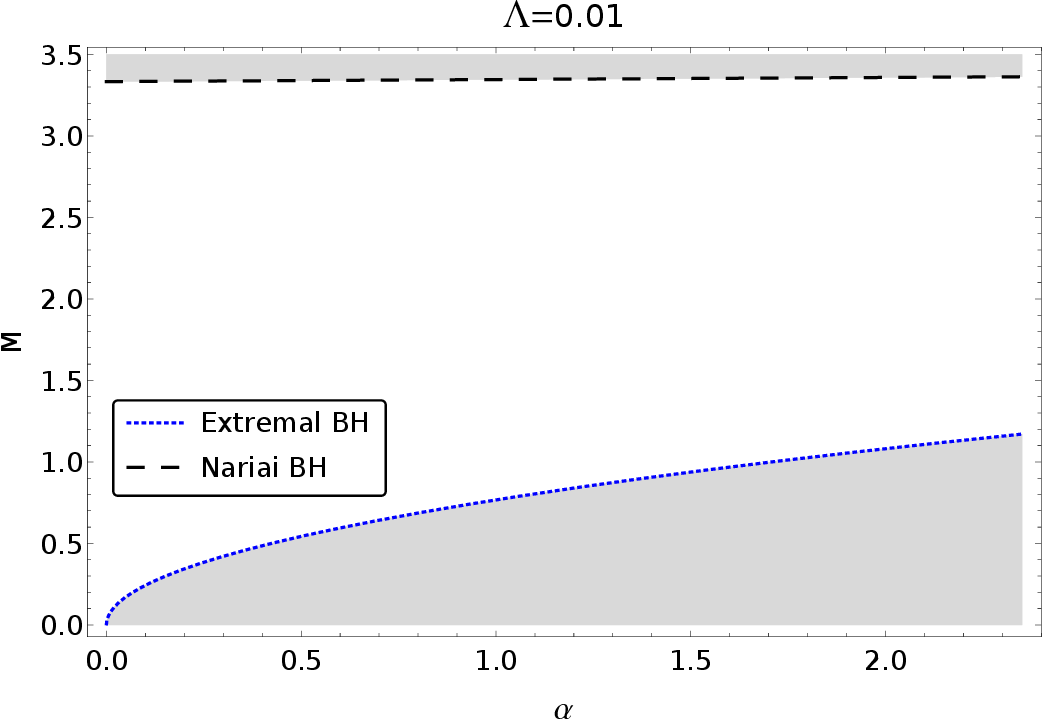}
\caption{Left: The metric function $f(r)$ versus $r$ obtained for given $\Lambda  = 0.1$ and $\alpha =1.1663$.
Right: The allowed parameter space $(\alpha, M)$ of the loop quantum corrected black hole in the white region for $\Lambda=0.01$.
}\label{Fig1}
\end{figure}

So far, we have built our model and sketched the calculation to get the metric of the outside region by the junction condition. 
Then, just following the procedure shown in \cite{Lewandowski:2022zce}, we get 
\begin{equation}\label{eq:metric}
f(r)=g(r)=1-\left(\frac{2 GM}{r}+\frac{\Lambda r^2}{3}-\frac{\alpha G^2 M^2}{r^4}\left(1+\frac{\Lambda r^3}{6 G M}\right)^2\right),
\end{equation}
where $\alpha=16\sqrt{3}\pi\gamma^3 G\hbar$, proportional to the Planck area, is the quantum deformation parameter. 
It should be noted that the metric \eqref{eq:metric} is valid only for $r>r_b$ with $r_b$ denoting the minimal radial of the dust ball at which the bounce occurs \cite{Lewandowski:2022zce}. 
In the left side of Fig. \ref{Fig1}, we plot the values of $f(r)$ depending on $r$ for $\Lambda=0.1$ and $\alpha =1.1663$, in which $M$ can take different values. 
As shown in the figure, for $M$ bigger than some extreme value $M_{\rm Ext}$, the metric function $f(r)$ has three roots, corresponding to the three horizons of the black hole.
They are respectively the Cauchy horizon ${r_i}$, the event horizon ${r_h}$ and the cosmological horizon ${r_c}$, with $r_i<r_h<r_c$.
If one may consider a process where the mass of the black hole increases for the given cosmological constant and loop quantum correction, the Nariai solutions take place as the cosmological horizon coincides with the event horizon, which is shown as a magenta line.
If one decreases the mass of the black hole, the Cauchy and the event horizons gradually approach.
When the Cauchy horizon coincides with the event horizon, the mass reaches an extreme value, which is denoted as $M_{\text{ext}}$. 
For $M<M_{\rm ext}$, the event horizon disappears resulting in a naked singularity. 
This case is thus prohibited by the weak cosmic censorship conjecture.
So the corresponding solution space is between the extremal solutions and the Nariai solutions.
As a demonstration, we restrict the black hole mass $M$ and the loop quantum correction $\alpha$ in the white region in the right plot of Fig. \ref{Fig1} for $\Lambda=0.01$, where the dotted blue line denotes extremal black hole and dashed black line corresponds to Nariai black hole. 
\section{\label{section3}The quasinormal modes for massless scalar perturbations}
Now, we consider a massless neutral scalar field perturbation in the above background.
The equation of motion in such a curved spacetime is governed by the following Klein-Gordon equation:
\begin{equation}\label{N8e}
\square \Phi  = 0.
\end{equation}

According to spherical symmetry of the spacetime, the scalar field can be expanded as
\begin{equation}\label{N8}
\Phi=\frac{\phi(r)}{r} Y_{lm}(\theta, \varphi) e^{-i \omega t},
\end{equation}
where ${Y_{lm}}(\theta ,\varphi )$ is the spherical harmonics function.
By plugging it into the Klein-Gordon equation, the master equation in the radial part reads
\begin{equation}\label{Nequa}
\left( {\frac{{{d^2}}}{{d{r_*^2}}} + {\omega ^2} - {V_{\rm eff}}(r)} \right)\phi (r) = 0,
\end{equation}
where the effective potential is given by
\begin{equation}\label{N10}
{V_{\rm eff}}(r) = f(r)\left[ {\frac{{l(l + 1)}}{{{r^2}}} + \frac{{f'(r)}}{r}} \right].
\end{equation}
$dr_*$ is tortoise coordinate, which is defined as $dr_*= \frac{dr}{f(r)}$.
Physically, there only exist purely ingoing waves near the event horizon and purely outgoing waves near the cosmological horizon~\cite{RevModPhys.83.793}. 
Thus the boundary conditions are imposed as 
\begin{equation}\label{conN11}
\phi (r) \approx {e^{ - i\omega {r_*}}}\;\left( {r \to {r_h}} \right),\quad \phi (r) \approx {e^{i\omega {r_*}}}\;\left( {r \to {r_c}} \right).
\end{equation}
Then, the discrete quasinormal frequencies can be derived by solving the equation of motion with the above boundary conditions ~(\ref{conN11}).

In the remainder of this paper, we will use three numerical methods to accurately calculate the lowest-lying quasinormal modes and present some relevant results.
Presently, many numerical computations of quasinormal modes have been developed with high precision~\cite{prd-qnm-anal-02,prs-qnm-anal-conti-01,Cho:2011sf}.
Here, we introduce the finite difference method~\cite{prd-qnm-lateti-Linear-01} to obtain the numerical evolution of the scalar field and then extract the quasinormal spectrum from the data samples with Prony method~\cite{prd_prony}.
In order to check the correctness of our results, we also employ the matrix method~\cite{agr-qnm-lq-matrix-02} and WKB approximation~\cite{aj-nm-semianal-wkb-01,prd-nm-wkb-02,prd-nm-wkb-03}.

First, it is necessary to perform a coordinate transformation to derive the double null coordinates, which is defined as $u = t - {r_*}$ and $v = t + {r_*}$.
Accordingly, the Klein-Gordon equation can be expressed as
\begin{equation}\label{N15}
 - 4\frac{{{\partial ^2}\phi }}{{\partial u\partial v}} = {V_{{\rm{eff}}}}(r(u,v))\phi.
\end{equation}
According to finite difference scheme, the data at $N$ can be obtained from $W$, $E$, and $S$, such that the above equation of motion gives rise to
\begin{equation}\label{N16}
{\phi _N} = {\phi _W} + {\phi _E} - {\phi _S} - \Delta u\Delta v{V_{{\rm{eff}}}}(r(u,v))\frac{{{\phi _W} + {\phi _E}}}{8},
\end{equation}
where the indices $N,W,E,S$ denote grid-points, respectively corresponding to the points $N \equiv (u + \Delta ,v + \Delta )$, $W \equiv (u,v + \Delta )$, $E \equiv (u + \Delta ,v)$, and $S \equiv (u,v)$ with $\Delta $ the step width of $(u,v)$. 
The time-domain profile will appear soon, once one provides the specific initial conditions 
\begin{equation}\label{N17}
\phi (u,0) = 0,\quad \phi (0,v) = {e^{ - \frac{{{{\left( {v - {v_c}} \right)}^2}}}{{2{\sigma ^2}}}}},
\end{equation}
where ${v_c}$ and $\sigma $ correspond to the center and width of the Gaussian wave packet.
The resulting temporal evolution $\phi (t,{r_*})$ can be obtained from equally elapsed late-time data.

Next, to extract the quasinormal mode from the temporal evolution data, Prony method is a very useful tool, which as an extension of the Fourier decomposition, is of great significance for signal processing and data analysis.
The late-time signal at a certain ${r_*}$ is composed of a set of quasinormal modes, which can be expanded as
\begin{equation}\label{N18}
\phi (t) = \sum\limits_{j = 1}^p {{C_j}} {e^{ - i{\omega _j}t}}. 
\end{equation} 
The time interval of the time-domain profile is between ${t_0}$ and $t={t_0} + qh $, where $h$ is the time interval of each point.
$q$ as the number of sample signals is an integer and satisfies $q= 2p$.
For convenience, every sample is labeled by an integer $n$.
According to the above formula, the time-domain data at any time can be expressed as
\begin{equation}\label{NX}
{x_n} = \sum\limits_{j = 1}^p {{{\tilde C}_j}} z_j^n,
\end{equation} 
where ${x_n} = \phi \left( {{t_0} + nh} \right),z_j = {e^{ - i{\omega _j}h}},{{\tilde C}_j} = {C_j}{e^{ - i\omega {t_0}}}$.
In order to find ${z_j}$, it is necessary to introduce a polynomial function
\begin{equation}\label{NAZ}
A(z) = \prod\limits_{j = 1}^p {\left( {z - {z_j}} \right)}  = \sum\limits_{i = 0}^p {{\alpha _i}} {z^{p-i}},
\end{equation} 
with ${\alpha _0} = 1$.
Obviously, for any integer $j$ from 1 to $p$, $A({z_j}) = 0$.
Thus, it's easy to obtain the sum
\begin{equation}\label{N21}
\sum\limits_{i = 0}^p {{\alpha _i}} {x_{j - i}} = \sum\limits_{i = 0}^p {{\alpha _i}} \sum\limits_{k = 1}^p {{{\tilde C}_k}} z_k^{j - i} = \sum\limits_{k = 1}^p {{{\tilde C}_k}} z_k^{j - p}A\left( {{z_k}} \right) = 0.
\end{equation} 
Considering ${\alpha _0} = 1$, the above equation can be rewritten as
\begin{equation}\label{N20}
\sum\limits_{i = 1}^p {{\alpha _i}} {x_{j - i}} =  - {x_j}.
\end{equation} 
Thus, we can get $p$ equations after taking $j$ from $p+1$ to $q$ such that ${\alpha _i}$ can be solved. 
After substituting ${\alpha _i}$ into Eq. (\ref{NAZ}), ${z_j}$ can be derived easily.
Then the quasinormal modes are obtained with the relation ${\omega _j} = \frac{i}{h}\ln \left( {{z_j}} \right)$.
The coefficients ${{{\tilde C}_j}}$ can also be found according to Eq. (\ref{NX}).

As a comparison, we further resort to the matrix method and WKB approximation to ensure the accuracy of numerical results.
For the matrix method, after the metric function $f(r)$ is expanded by Taylor series near the event and cosmological horizons, the boundary conditions can be rewritten as
\begin{equation}\label{N21}
\phi \left( {{r_h}} \right) \approx {\left( {r - {r_h}} \right)^{ - \frac{{i\omega }}{{f'\left( {{r_h}} \right)}}}},\quad \phi \left( {{r_c}} \right) \approx {\left( {{r_c} - r} \right)^{\frac{{i\omega }}{{f'\left( {{r_c}} \right)}}}}.
\end{equation} 
More specifically, the wave function of the scalar field $\phi (r)$ can be transformed into $Y(y)$ through the following relationship
\begin{equation}\label{N19}
\phi (r)  = {y^{ - \frac{{i\omega }}{{{f^\prime }\left( {{r_h}} \right)}}-1}}{(1 - y)^{\frac{{i\omega }}{{{f^\prime }\left( {{r_c}} \right)}}-1}}Y (y),
\end{equation}
where  $y = \frac{{r - {r_h}}}{{{r_c} - {r_h}}}$.
Correspondingly, the perturbation equations and the desired boundary conditions reduce to
\begin{equation}\label{N20}
\begin{array}{l}
{b_0}(\omega ,y)Y(y) + {b_1}(\omega ,y){Y^\prime }(y) + {b_2}(\omega ,y){Y^{\prime \prime }}(y) = 0,\\
Y\left( 0 \right) = Y\left( 1 \right) = 0.
\end{array}
\end{equation}
By discretizing the above equation, an equation with the matrix form can be expressed as $\Gamma(\omega)\mathcal{Y}  = 0$.
The quasinormal modes can be determined by solving the nonlinear algebraic equation $\det (\Gamma (\omega )) = 0$.

The WKB approximation is also a semianalytic approach to calculate the quasinormal modes.
The quasinormal modes can be derived by the six-order WKB formula, which reads 
\begin{equation}\label{N21}
\frac{i\left(\omega^2-{{V_{\rm eff}}}\left(r_0\right)\right)}{\sqrt{-2 {{V_{\rm eff}}}^{\prime \prime}\left(r_0\right)}}+\sum_{j=2}^6 \Lambda_j=n+\frac{1}{2},
\end{equation}
where the prime $'$ represents the derivative with respect to tortoise coordinate ${r_*}$ and ${{r_0}}$ is the maximum extremum of the effective potential.
$n$ is the overtone and ${{\Lambda _j}}$ is the higher order correction term that can be found in~\cite{article,PhysRevD.35.3621,PhysRevD.35.3632,PhysRevD.68.024018}. 

\begin{table}[]
\caption{The lowest-lying quasinormal modes of $l=1$ with different modes, where the results are obtained by different numerical methods for $M=2$ and $\Lambda  = 0.01$.
}
\setlength{\tabcolsep}{9.3mm}
\begin{tabular}{|cccc|}
\hline
\multicolumn{4}{|c|}{$l=1$}                                                                                                         \\ \hline
\multicolumn{1}{|c|}{$\alpha$} & \multicolumn{1}{c|}{Prony}            & \multicolumn{1}{c|}{Matrix method}                 & WKB approximation                \\ \hline
\multicolumn{1}{|c|}{0}    & \multicolumn{1}{c|}{0.11229-0.04101i} & \multicolumn{1}{c|}{0.11234- 0.04103i}  & 0.11234 - 0.04104i \\ \hline
\multicolumn{1}{|c|}{0.08706}  & \multicolumn{1}{c|}{0.11240-0.04099i} & \multicolumn{1}{c|}{0.11246 - 0.04101i} & 0.11246 - 0.04103i \\ \hline
\multicolumn{1}{|c|}{0.69650}  & \multicolumn{1}{c|}{0.11323-0.04083i} & \multicolumn{1}{c|}{0.11327 - 0.04085i} & 0.11326 - 0.04089i \\ \hline
\multicolumn{1}{|c|}{2.35068}  & \multicolumn{1}{c|}{0.11551-0.04029i} & \multicolumn{1}{c|}{0.11557 - 0.04029i} & 0.11557 - 0.04035i \\ \hline
\end{tabular}
\label{tab1}
\end{table}

\begin{table}[]
\caption{The lowest-lying quasinormal modes of $l=2$ with different modes, where the results are obtained by different numerical methods for $M=2$ and $\Lambda  = 0.01$.
}
\setlength{\tabcolsep}{9.3mm}
\begin{tabular}{|cccc|}
\hline
\multicolumn{4}{|c|}{$l=2$}                                                                                                         \\ \hline
\multicolumn{1}{|c|}{$\alpha$} & \multicolumn{1}{c|}{Prony}            & \multicolumn{1}{c|}{Matrix method}                 & WKB approximation                  \\ \hline
\multicolumn{1}{|c|}{0}    & \multicolumn{1}{c|}{0.19039-0.03935i} & \multicolumn{1}{c|}{0.19039 - 0.03938i} & 0.19039 - 0.03938i \\ \hline
\multicolumn{1}{|c|}{0.08706}  & \multicolumn{1}{c|}{0.19057-0.03933i} & \multicolumn{1}{c|}{0.19057 - 0.03936i} & 0.19057 - 0.03937i \\ \hline
\multicolumn{1}{|c|}{0.69650}  & \multicolumn{1}{c|}{0.19182-0.03921i} & \multicolumn{1}{c|}{0.19182 - 0.03924i} & 0.19182 - 0.03925i \\ \hline
\multicolumn{1}{|c|}{2.35068}  & \multicolumn{1}{c|}{0.19539-0.03876i} & \multicolumn{1}{c|}{0.19540 - 0.03879i} & 0.19540 - 0.03880i \\ \hline
\end{tabular}
\label{tab2}
\end{table}

\begin{figure}[H]
\centering
\includegraphics[scale=0.48]{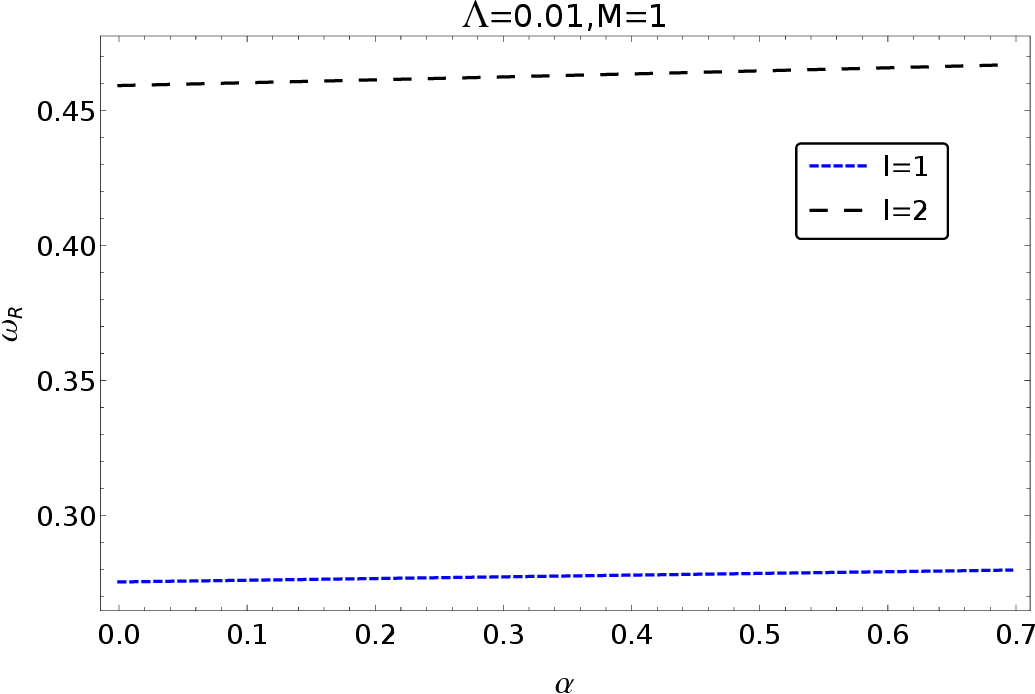}
\includegraphics[scale=0.5]{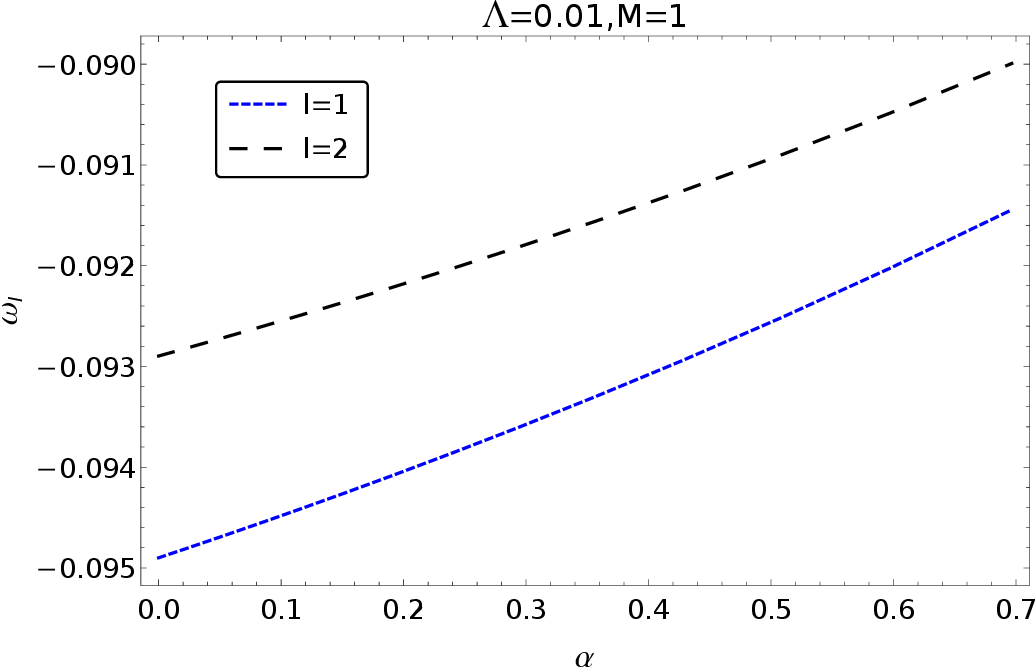}
\caption{The lowest-lying quasinormal modes as a function of the loop quantum correction $\alpha$ with $l=1$ and $l=2$, where $M=1$ and $\Lambda  = 0.01$ are set.
}\label{Fig2}
\end{figure}

In Tab.\ref{tab1} and \ref{tab2}, we present low-lying quasinormal modes for the massless neutral scalar field obtained from Prony method, matrix method, and WKB approximation.
As shown in Tab.\ref{tab1} and \ref{tab2}, the numerical results derived by all methods are consistent with each other, and their accuracy error is controlled within 5 percent, which demonstrates the reliability of our numerical calculations.
Our results also indicate that the loop quantum correction $\alpha$ has a significant impact on the quasinormal modes.
For both $l=1$ and $l=2$ modes, as $\alpha$ increases, the real part of the quasinormal frequency increases, which implies that $\alpha$ can amplify the oscillation frequencies of waveforms.
By contrast, the magnitude of the imaginary part decreases with an increase of loop quantum correction $\alpha$, indicating a slower dissipation of the massless neutral scalar field.
As evident from Fig.\ref{Fig2}, we present the lowest-lying quasinormal modes by varying the loop quantum correction $\alpha$, which further highlights the observation that the quantum correction can increase the real part of the quasinormal modes and decrease the magnitude of the imaginary part.
Moreover, the absence of non-negativity in the imaginary part of quasinormal modes indicates this quantum-modified black hole is stable against the scalar perturbations.
 
\begin{figure}[htbp]
\centering
\includegraphics[scale=0.48]{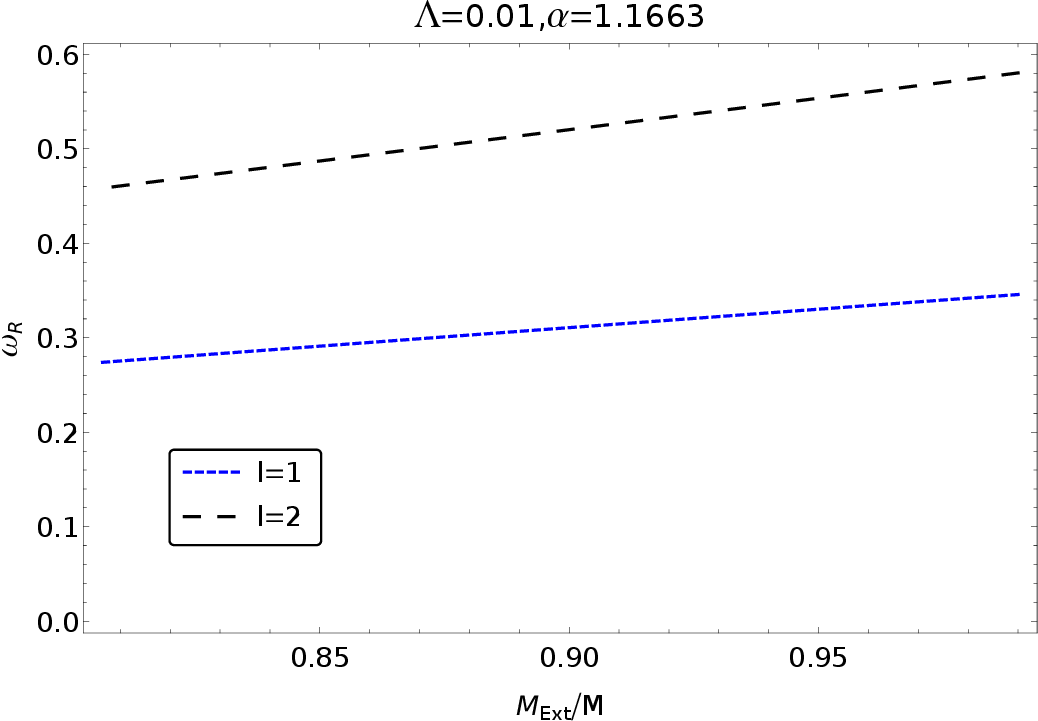}
\includegraphics[scale=0.5]{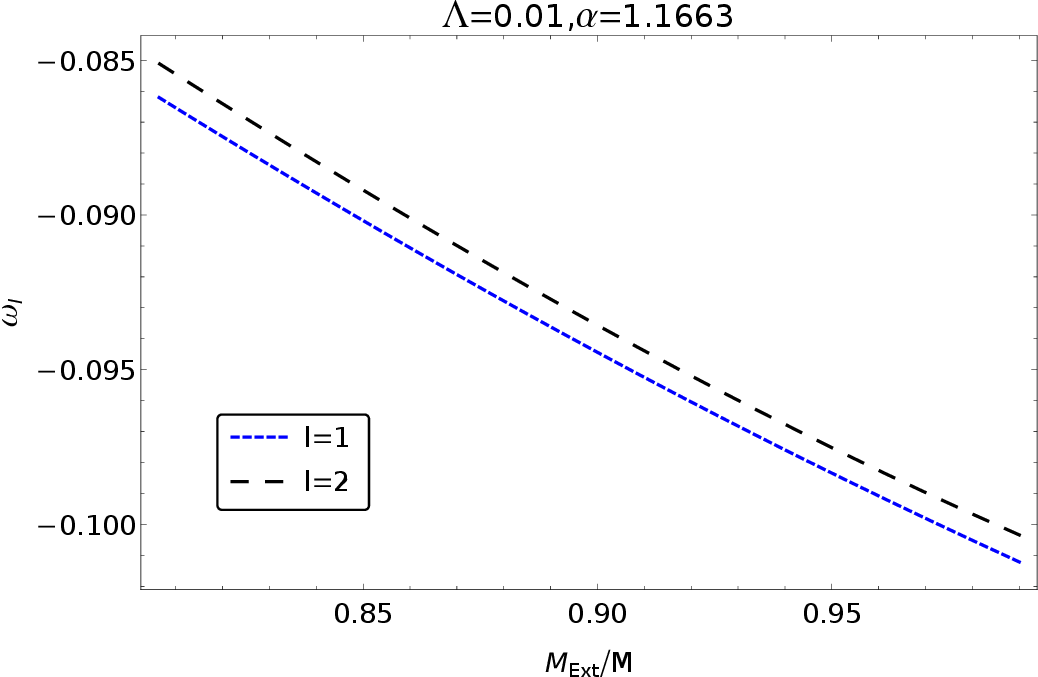}\vskip 15pt
\includegraphics[scale=0.48]{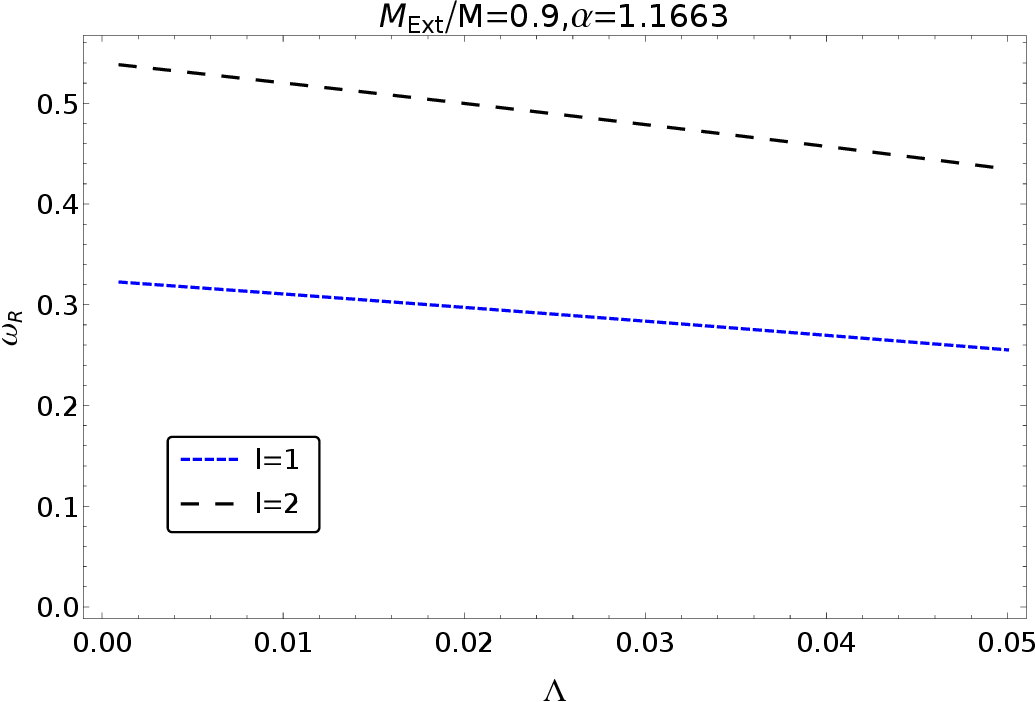}
\includegraphics[scale=0.5]{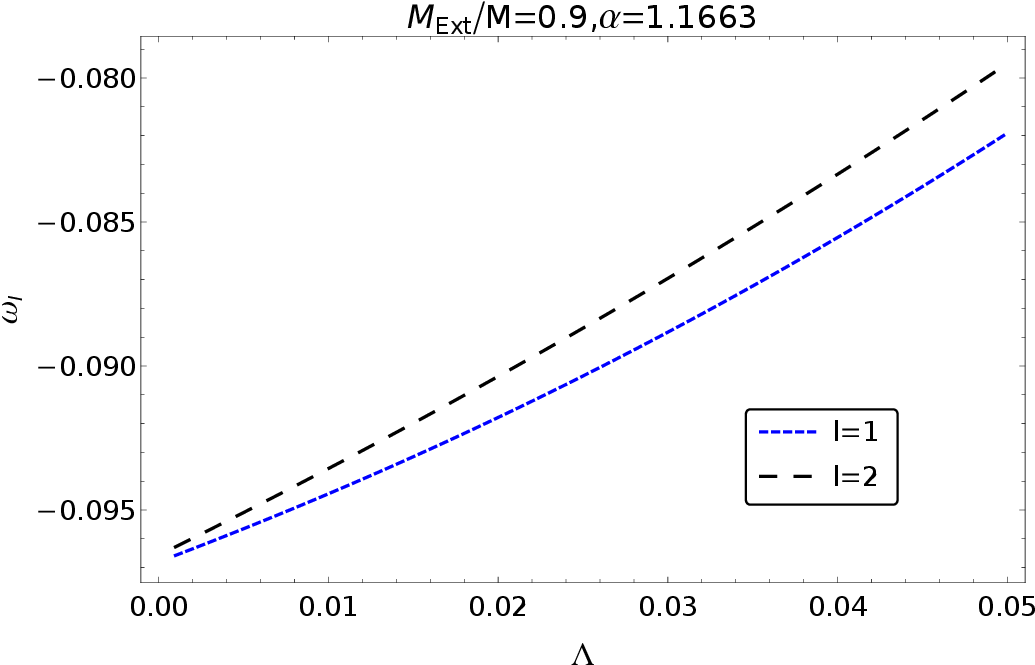}
\caption{Top: The lowest-lying quasinormal modes as a function of the black hole mass ratio ${M_{E{\rm{xt}}}}/M$ with $l=1$ and $l=2$, where $\alpha=1.1663$ and $\Lambda  = 0.01$ are set.
Bottom: The lowest-lying quasinormal modes as a function of the cosmological constant $\Lambda$ with $l=1$ and $l=2$, where ${M_{E{\rm{xt}}}}/M = 0.9$ and $\alpha=1.1663$ are set.
}\label{Fig3}
\end{figure}

In addition, we also explore the impact of the black hole mass ratio ${M_{E{\rm{xt}}}}/M$ and the cosmological constant $\Lambda$ on the lowest-lying quasinormal modes, as shown in Fig.\ref{Fig3}.
One observes that both the real part and the magnitude of the imaginary part of quasinormal frequency grow with an increase of the black hole mass ratio ${M_{E{\rm{xt}}}}/M$. 
This implies as the black hole approaches extremality, the period of the oscillation decreases and decays faster in time.
On the contrary, as the cosmological constant $\Lambda$ increases, both the oscillation and decay of the scalar field become slower.
After comparison, it is found that a non-vanishing $\alpha$ can make $\Lambda$ have a weaker effect on the decay of the field than a vanishing $\alpha$.
\section{\label{section4}Late-time tails}
In this section, we further explore the reactions of the loop quantum corrected black hole under scalar perturbations and investigate how the asymptotic behaviors of the late-time tail can be affected by the cosmological constant, the multipole number, the black hole mass ratio, and the loop quantum correction.
\begin{figure}[htbp]
\centering
\includegraphics[scale=0.5]{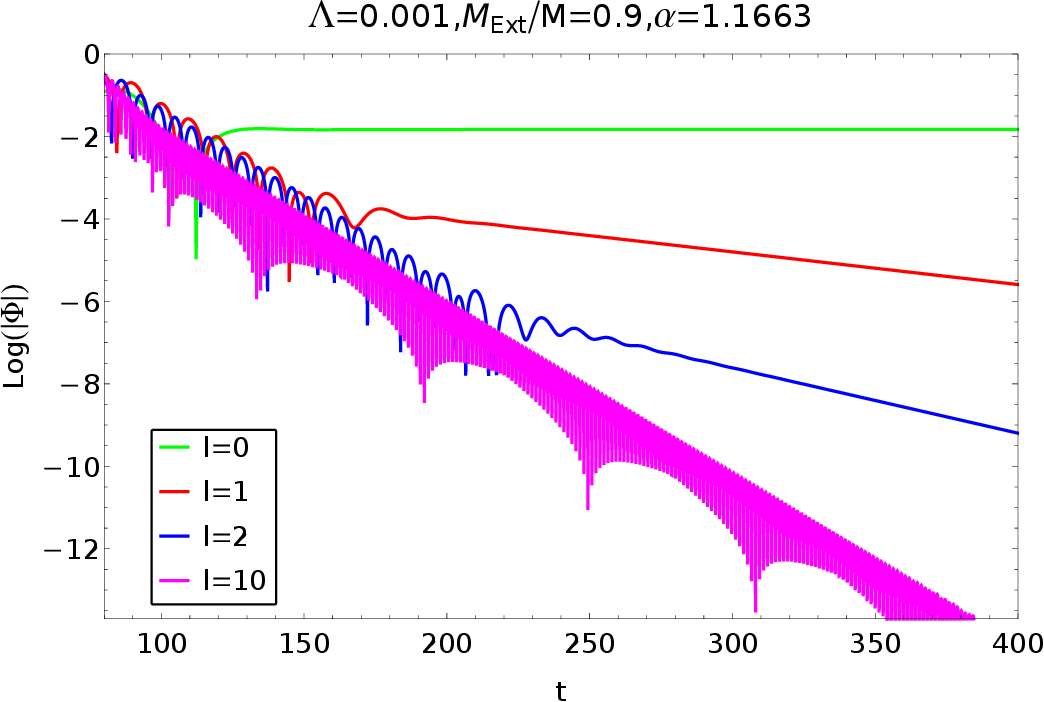}
\includegraphics[scale=0.5]{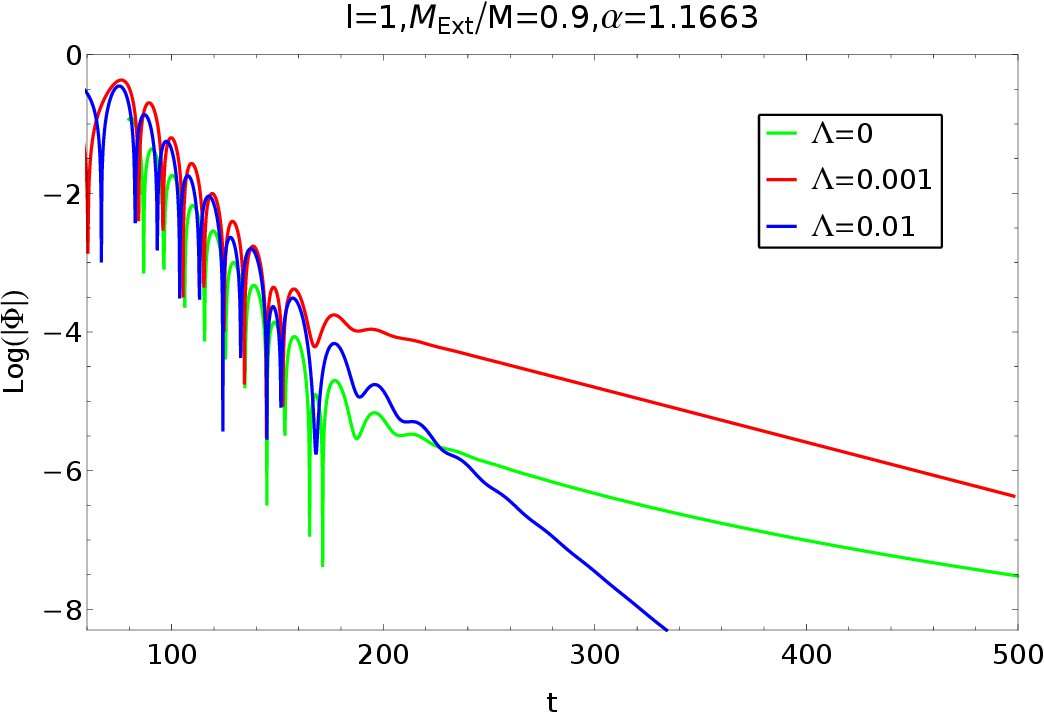}\vskip 15pt
\includegraphics[scale=0.5]{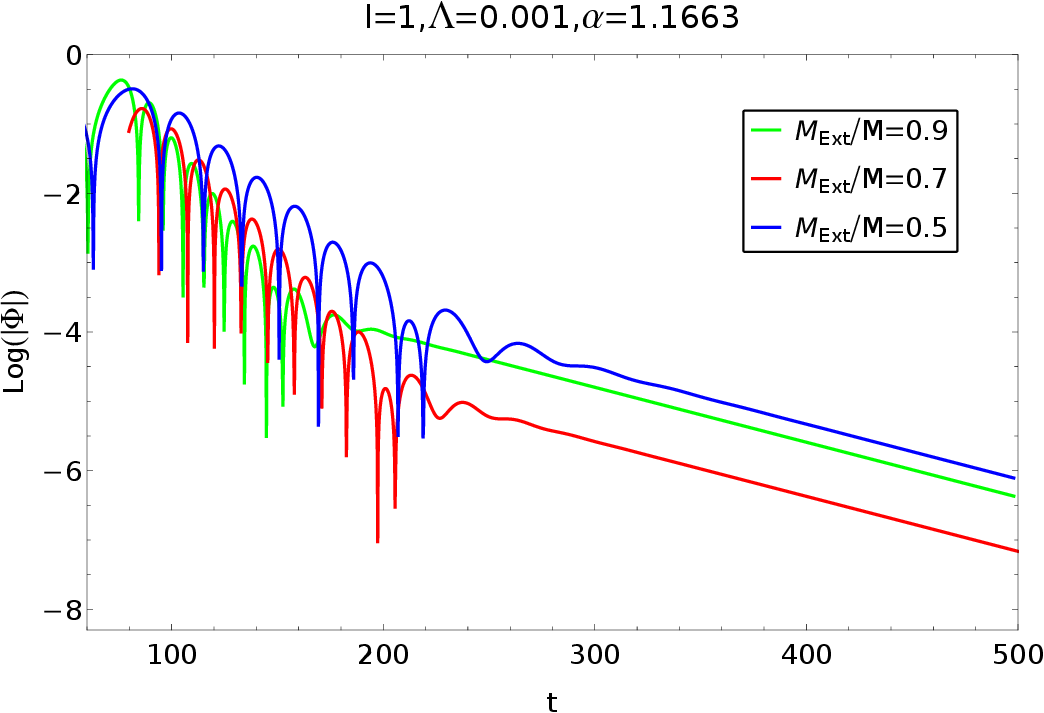}
\includegraphics[scale=0.5]{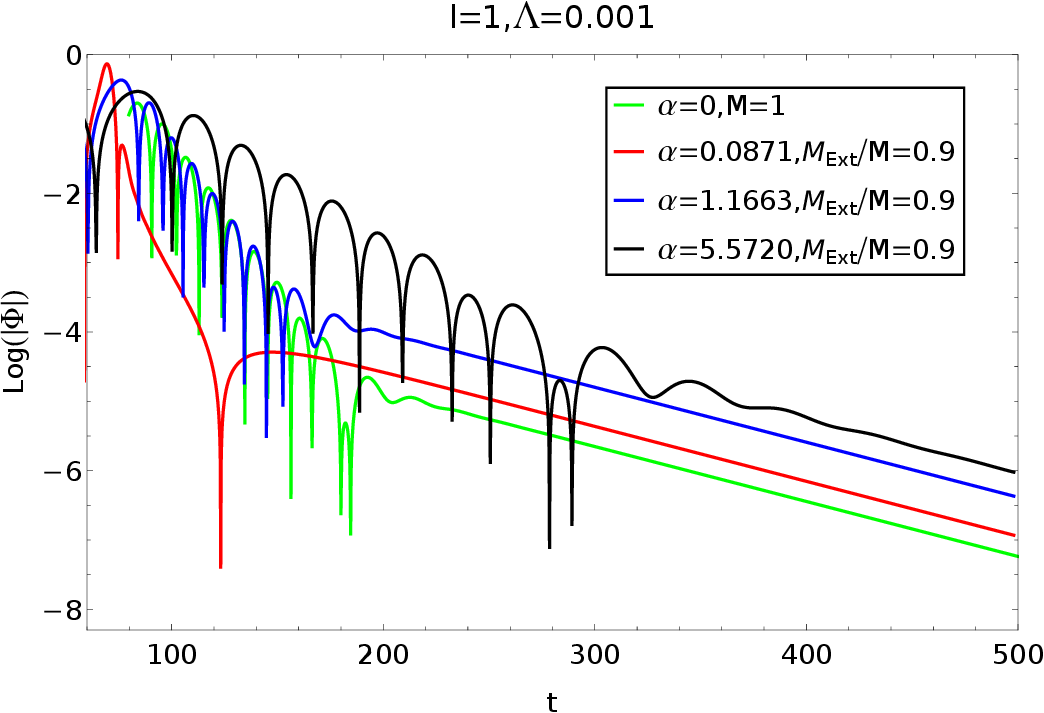}
\caption{Top left: The temporal evolution of massless scalar perturbations with ${M_{E{\rm{xt}}}}/M = 0.9$, $\Lambda  = 0.001$, and $\alpha=1.1663$ on the multipole number $l=0$, $l=1$, $l=2$, and $l=10$.
Top right: The temporal evolution of massless scalar perturbations with $l=1$, ${M_{E{\rm{xt}}}}/M = 0.9$, and $\alpha=1.1663$ on the cosmological constant $\Lambda  = 0$, $\Lambda  = 0.001$ and $\Lambda  = 0.01$.
Bottom left: The temporal evolution of massless scalar perturbations with $l=1$, $\Lambda  = 0.001$, and $\alpha=1.1663$ on the black hole mass ratio ${M_{E{\rm{xt}}}}/M = 0.9$, ${M_{E{\rm{xt}}}}/M = 0.7$, and ${M_{E{\rm{xt}}}}/M = 0.5$.
Bottom right: The temporal evolution of massless scalar perturbations with $l=1$ and $\Lambda  = 0.001$ on different loop quantum corrections.
}\label{Fig4}
\end{figure}
The late-time tail dominates the last stage of the temporal evolution of the scalar field.
Past studies~\cite{Leaver:1986gd,Ching:1995tj,Leaver:1986vnb} indicate that the quasinormal frequencies correspond to the poles of Green's function, while the late-time tail is mathematically governed by branch cuts of Green's function in the complex frequency plane.
From a physical perspective, researchers have thought the late-time tails are mostly determined by the asymptotics of the effective potential at spatial infinity~\cite{prd-qnm-lateti-Linear-01,Ching:1995tj}.
In asymptotically flat spacetimes, since the backscattering of perturbed wave packets from asymptotically far regions, inverse power-law tails are mostly observed~\cite{Ching:1995tj,Koyama:2000hj,Yu:2002st,Poisson:2002jz}.
As expected, if we remove the cosmological constant in such a model, the radiative decay follows inverse power-law form as shown in the top right of Fig.\ref{Fig4}, where we present temporal evolutions of massless scalar perturbation, simulated by Gaussian wave packets.
On the other hand, in asymptotically de Sitter spacetime, the late-time tail becomes exponential.
As depicted in the figure, the late-time behavior is transformed from power-law decay to exponential falloff in the presence of a tiny positive cosmological constant, which is the same as that in RNdS spacetime~\cite{prd_ds_field}.
Currently, there is no exact analytical formula available for the late-time behavior in de Sitter spacetime.
Among the earlier efforts, it was found that the qualitative description of exponential decay law can be approximately expressed as~\cite{prd_sds_tail,prd_ds_field}
\begin{equation}\label{N21}
\phi  \approx {e^{ - \alpha \left( {l,s,{\kappa _c}} \right)t}},
\end{equation} 
where $l$ is the multipole number. $s$ represents the spin of fields and ${\kappa _c}$ is the surface gravity at the cosmological horizon.
Therefore, here we will discuss how different parameters in the model affect late-time behavior in de Sitter spacetime.
In the top left of Fig.\ref{Fig4}, we can see when $l=0$, the scalar perturbations decay to a non-zero constant at late times.
As the multipole number $l$ increases, the late-time tail decays faster.
These characteristics indicate the exponential tail can easily overpower the quasinormal frequency and purely imaginary quasinormal modes show up eventually.
Nevertheless, for propagating scalar field with larger $l$, instead of a tail, what finally emerges are the longest-lived quasinormal frequencies, which are prolonged forever.
Additionally, one observes that the larger cosmological constant can make the late-time tail decay faster in the top right of Fig.\ref{Fig3}.
In the bottom left of Fig.\ref{Fig3}, if we change the black hole mass ratio, the curves of late-time tails are essentially parallel, which implies the impact of the black hole mass ratio on the late-time behavior is relatively small.
As the loop quantum correction $\alpha$ fades away, this black hole can naturally return to Schwarzschild-de Sitter spacetime.
Due to the minor impact of the black hole mass on the late-time tail, we fix $M=1$ in Schwarzschild-de Sitter spacetime and fix ${M_{E{\rm{xt}}}}/M = 0.9$ in other cases to study the effect of loop quantum correction on the late-time tail.
In the bottom right of Fig.\ref{Fig4}, it is found the loop quantum correction plays the same role as the black hole mass ratio and the effect of loop quantum correction is insignificant for late-time tail.
Even if the loop quantum correction disappears, it does not have a significant impact on the behavior of the late-time tail.
\section{\label{section5} Validity of strong cosmic censorship}
Based on the analysis above, the presence of a cosmological constant can lead to an exponential decay of perturbations, at the same time, there is an exponential blue-shift amplification along the Cauchy horizon~\cite{Poisson:1990eh}.
The alarming fact is that the redshift effect of the remnant field can be enough to offset the blueshift effect, which determines the fate of SCC.
Therefore, it is interesting to check the validity of SCC for loop quantum gravity black holes in de Sitter spacetime.

Let us assume that the effective quantum action for loop quantum gravity with a quantum scalar field $\Phi $ can be written as:
\begin{equation}\label{N11aa}
S = \int_{} {{d^4}} x\sqrt { - g} \left( {\frac{{{\cal R} - 2\Lambda }}{{16\pi }} + {L_\hbar }(g) - \frac{1}{2}{\nabla _\mu }\Phi {\nabla ^\mu }\Phi  + {L_\hbar }(\Phi )} \right),
\end{equation}
where ${\cal R}$ is the Ricci scalar, ${{L_\hbar }(g)}$ is quantum corrections related to spacetime and ${{L_\hbar }(\Phi )}$ is the quantum correction part of the scalar field.
In semiclassical analysis, Hollands disregarded the term "${{L_\hbar }(g)}$" in equation (\ref{N11aa}) and investigated the impact of quantum effects within the quantum scalar field on the SCC in the framework of classical metric tensor~\cite{Hollands:2019whz}.
In contrast to Hollands' approach, we, in our semiclassical analysis, discard the term "${{L_\hbar }(\Phi )}$" in equation (\ref{N11aa}), allowing us to examine the validity of SCC by considering classical scalar fields in quantum-corrected spacetime.
To this end, we can compare the quantum effects in black hole spacetime with that in scalar field~\cite{Hollands:2019whz} to discover which one has a more significant impact on the SCC.
Furthermore, we are going to explore the effects of the black hole mass ratio, the cosmological constant on the validity of SCC for loop quantum gravity black holes under scalar field perturbation.
For research purposes, our focus here is only on a black hole with three horizons.

Now let us consider the relationship between quasinormal frequencies and the SCC.
If one imposes purely ingoing wave near the event horizon, the solution from the equation of motion has both the outgoing and ingoing waves near the Cauchy horizon, which can be expressed as 
\begin{equation}\label{N12}
{\phi _{\rm in}} \approx {e^{ - i\omega u}}{\left( {r - {r_i}} \right)^{\frac{{i\omega }}{{{\kappa _i}}}}}
,\quad \phi _{\rm out} \approx e^{ - i\omega u},
\end{equation}
where $u$ is outgoing coordinate defined as $u = t - r_*$ and ${\kappa _i}$ is the surface gravity of Cauchy horizon defined as ${\kappa _i} = \left| {\frac{1}{2}{f^\prime }\left( {{r_i}} \right)} \right|$.
Obviously, the ingoing wave has non-smooth radial dependence, which results in the potential non-smoothness behavior in the energy-momentum tensor of the scalar field. 
Commonly, the violation of the SCC implies the weak solution can be extended beyond the Cauchy horizon.
In other words, the energy-momentum tensor consisting of the square of its first derivative for the scalar field can be integrable at the Cauchy horizon, which requires~\cite{prl_scc_qnm}
\begin{equation}\label{N14}
\beta  =  - \frac{{{\mathop{\rm Im}\nolimits} \omega }}{{{\kappa _i}}} > \frac{1}{2}.
\end{equation}
for all the quasinormal modes.
On the contrary, as long as one finds the lowest lying quasinormal modes with the criterion $\beta  \le \frac{1}{2}$, the SCC is preserved.
Hence, in order to check the validity of the SCC, we exclusively focus on the lowest-lying quasinormal modes.
Note that there are three distinct families to classify the relevant quasinormal modes, namely, the near-extremal modes with $l=0$, the de Sitter modes with $l=1$, and the photon sphere modes with large $l$ s.
In what follows, we are going to check the validity of the SCC with these three modes.
\begin{figure*}[htbp]
\centering
\includegraphics[scale=0.33]{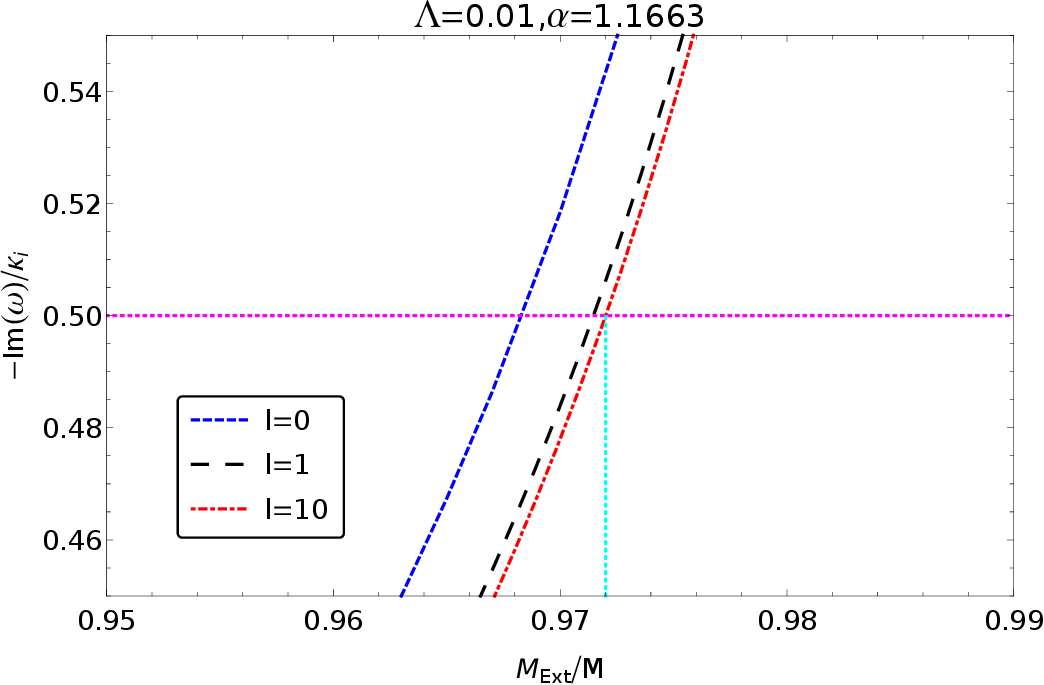}
\includegraphics[scale=0.33]{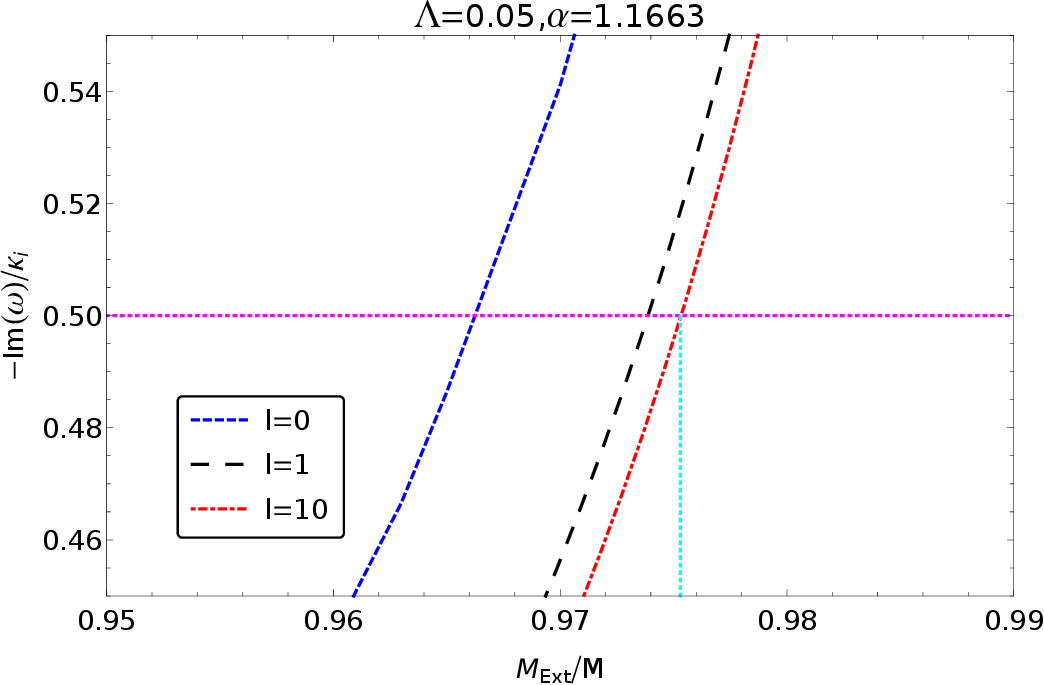}
\includegraphics[scale=0.33]{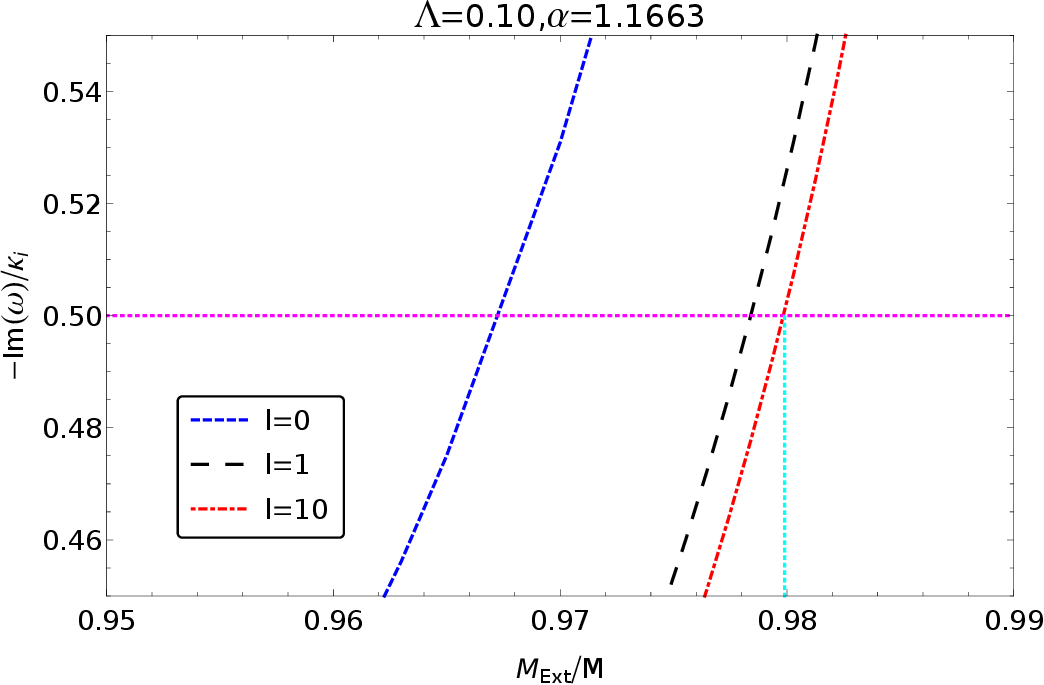}
\caption{ The lowest-lying quasinormal modes with the frequency $\beta=\frac{{ - {\mathop{\rm Im}\nolimits} (\omega )}}{{{\kappa _i}}}$ as a function of the black hole mass ratio ${M_{E{\rm{xt}}}}/M$, where the dotted magenta horizontal line represents the threshold value $\beta  = \frac{1}{2}$ and the dotted cyan vertical line denotes the critical value of the mass ratio for the violation of the SCC.
}\label{Fig5}
\end{figure*}

\begin{figure*}[htbp]
\centering
\includegraphics[scale=0.33]{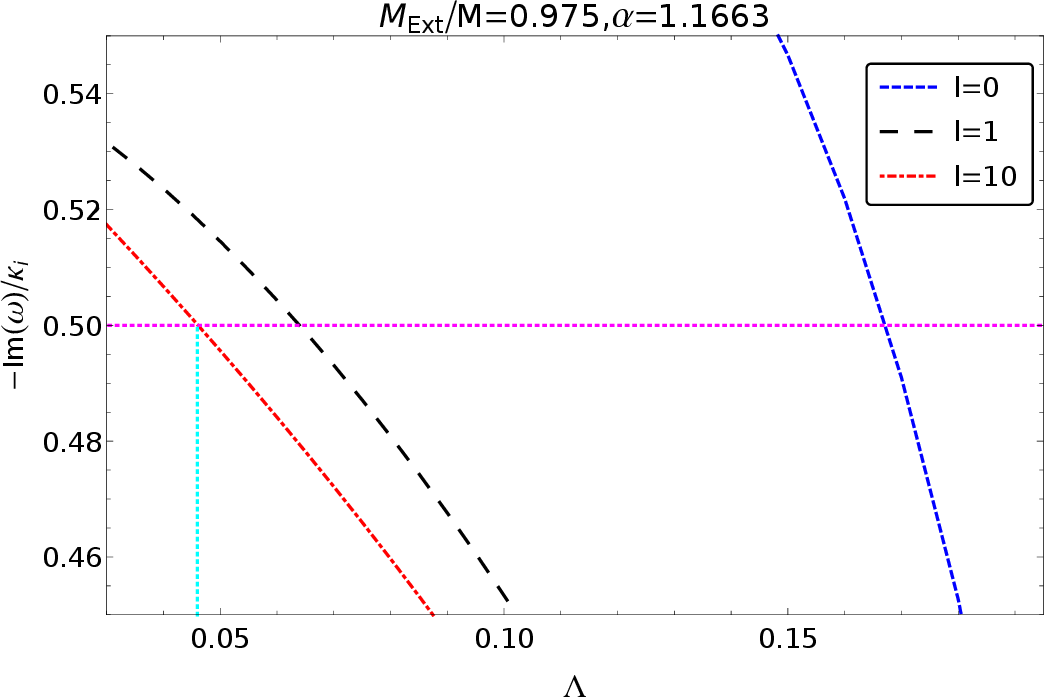}
\includegraphics[scale=0.33]{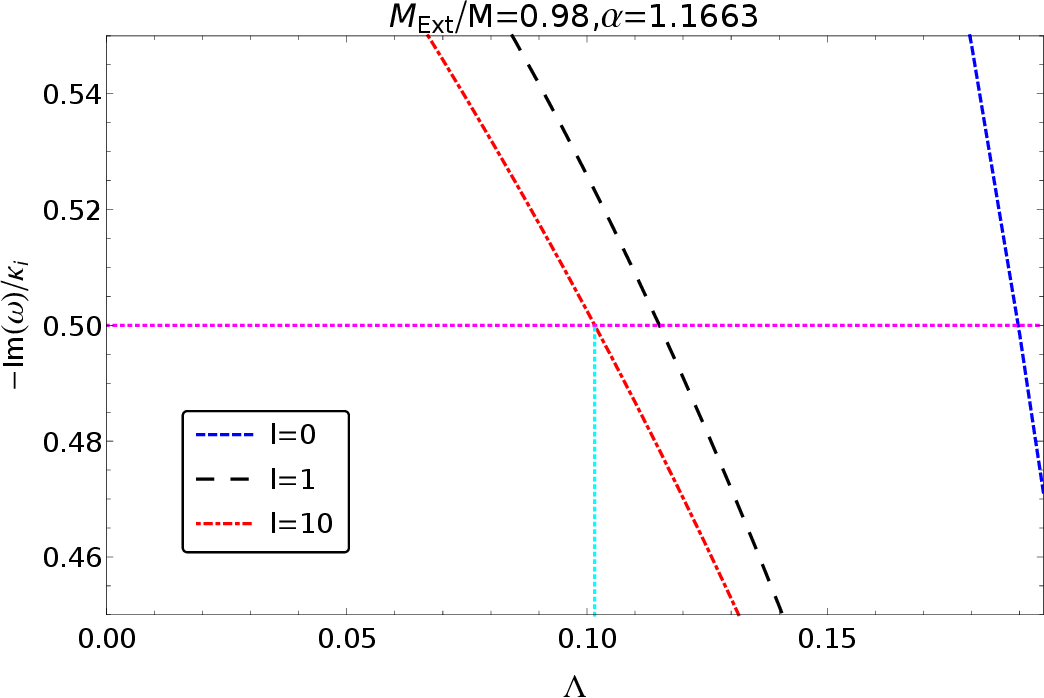}
\includegraphics[scale=0.33]{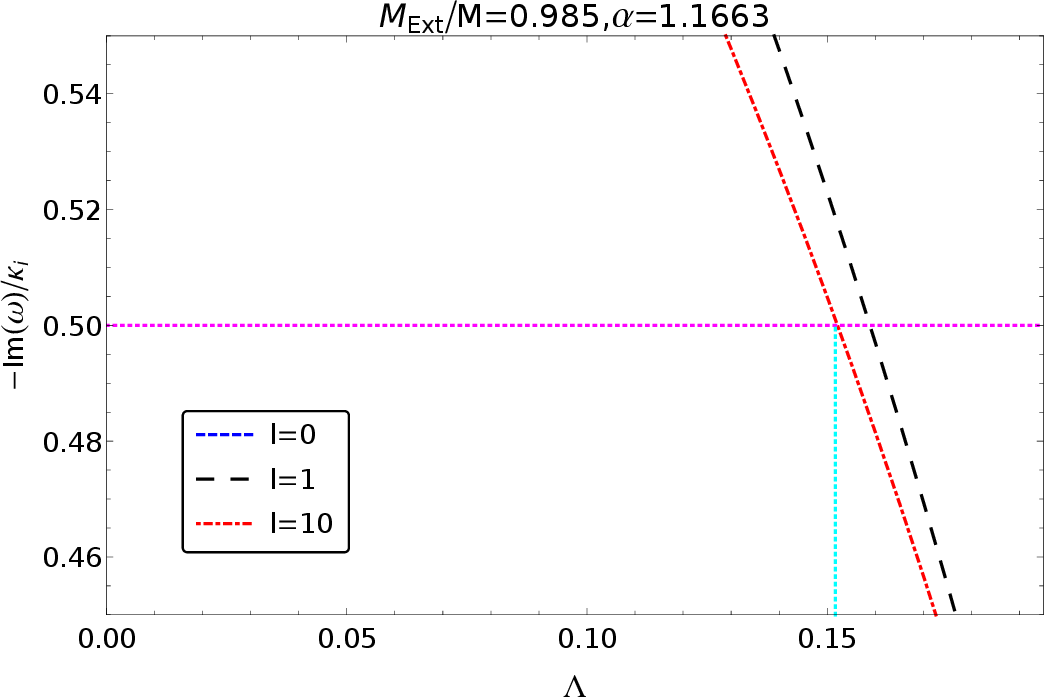}
\caption{The lowest-lying quasinormal modes with the frequency $\beta=\frac{{ - {\mathop{\rm Im}\nolimits} (\omega )}}{{{\kappa _i}}}$ as a function of the cosmological constant $\Lambda$, where the dotted magenta horizontal line represents the threshold value $\beta  = \frac{1}{2}$ and the dotted cyan vertical line denotes the critical value of $\Lambda$ for the restoration of the SCC.
}\label{Fig6}
\end{figure*}

As shown in Fig.\ref{Fig5}, we present the variation of $\beta$ with the black hole mass ratio ${M_{E{\rm{xt}}}}/M$ for different cosmological constants for the given $l$.
When the cosmological constant is fixed, $\beta$ becomes larger and larger as the mass of the black hole is close to the extremal limit.
It was implied that the SCC will only be violated as the mass ratio exceeds a certain critical value.
In addition, the threshold value of the mass ratio for the violation of the SCC increases with the cosmological constant. 
As a demonstration, we also plot the variation of $\beta$ with the cosmological constant $\Lambda$ for different black hole mass ratios ${M_{E{\rm{xt}}}}/M$ in Fig.\ref{Fig6}.
It was clear that the larger the cosmological constant is, the harder the SCC is violated.
This indicates that the cosmological constant plays an important role in recovering the SCC.
Moreover, the critical value for $\Lambda$ to rescue the SCC becomes larger with the increase of the mass ratio. 

\begin{figure*}[!]
\centering
\includegraphics[scale=0.5]{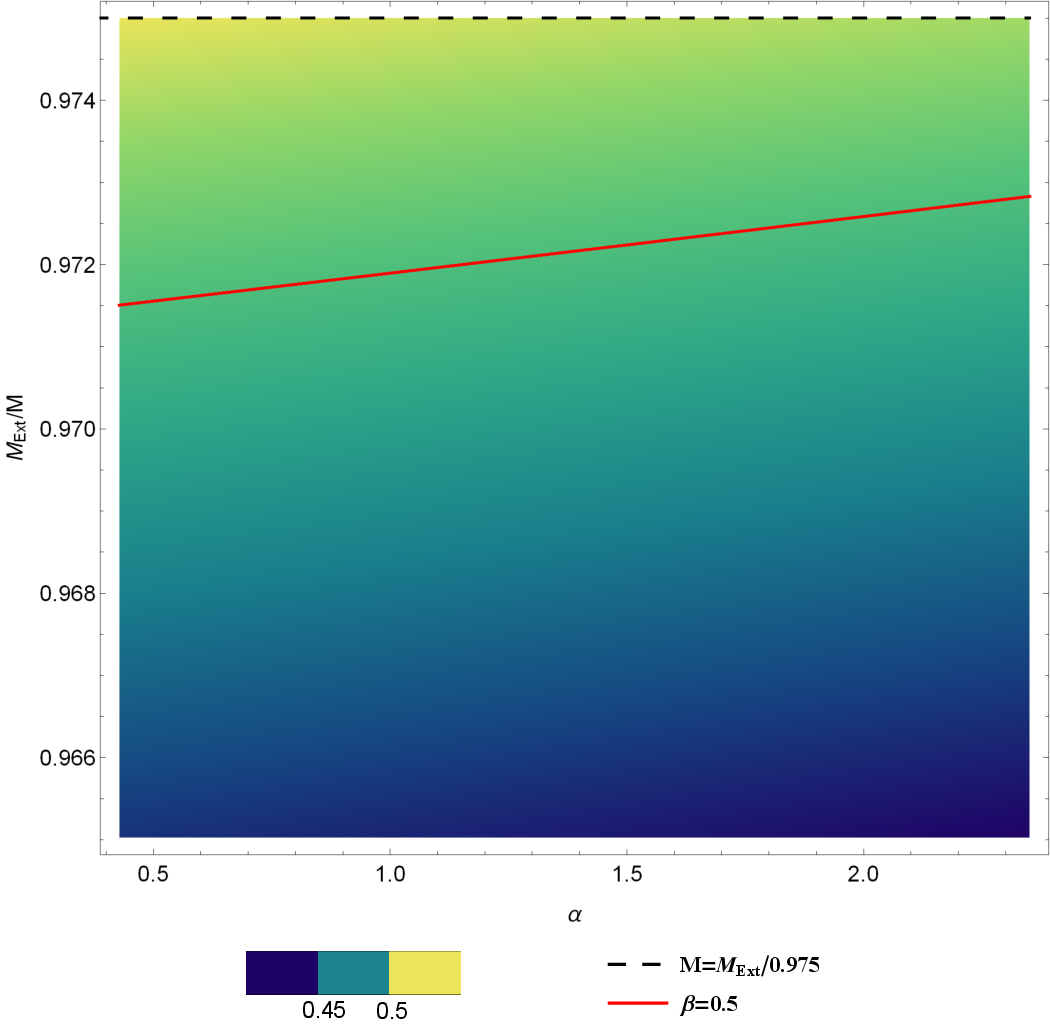}
\includegraphics[scale=0.5]{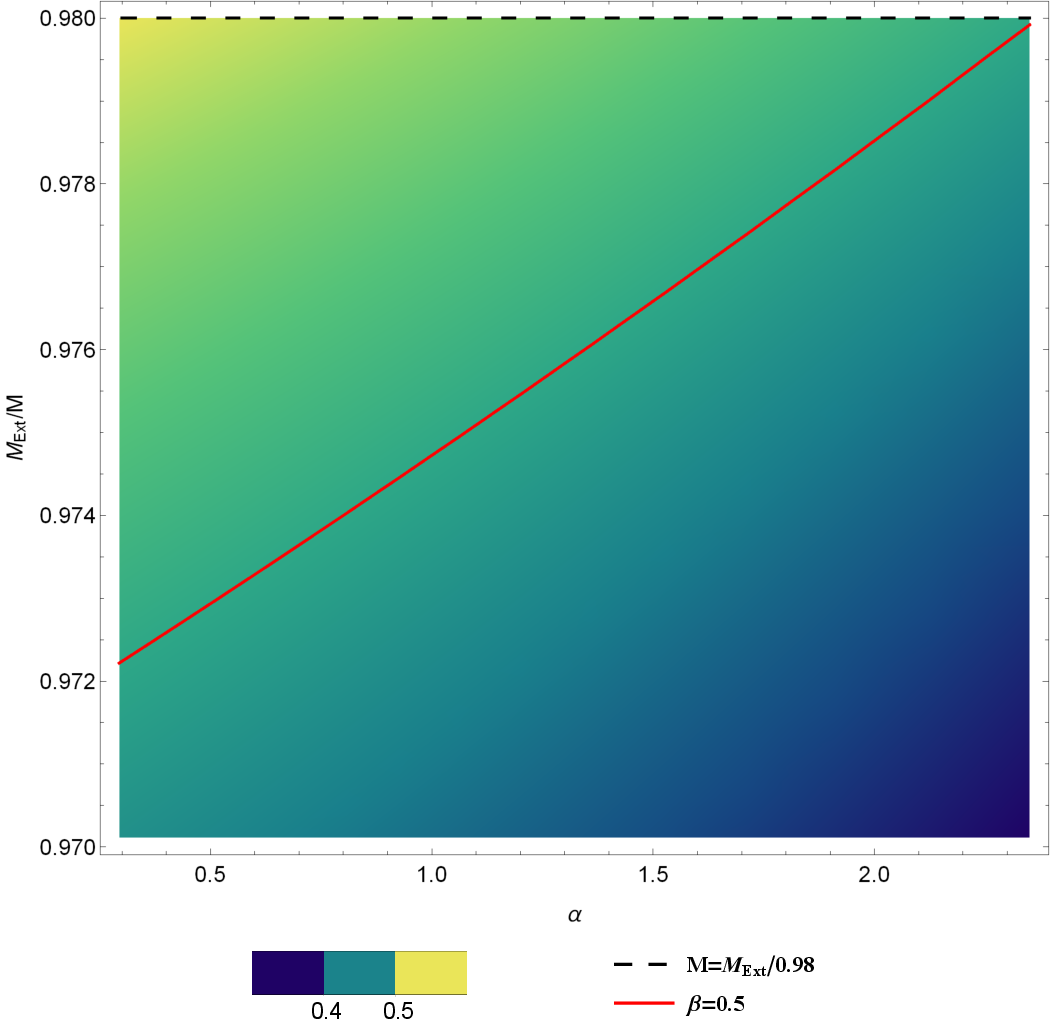}
\caption{The density plots of $\beta$ in the $\frac{{{M_{Ext}}}}{M} - \alpha $ plane with $\Lambda  = 0.01$ (left plane) and $\Lambda  = 0.05$ (right plane), where the region above the red line indicates the violation of the SCC.
}\label{Fig7}
\end{figure*}

Finally, to display the impact of loop quantum correction on the SCC, we present the density plots of $\beta$ in the $\frac{{{M_{Ext}}}}{M} - \alpha $ plane for $\Lambda  = 0.01$ and $\Lambda  = 0.05$ in Fig.\ref{Fig7}.
The critical threshold $\beta=1/2$ is marked as a solid red line.
Only in the area above the solid red line can the SCC be violated.
As expected, as long as the black hole approaches the extremal limit, the SCC is always violated.
As one can see, as the loop quantum correction increases, the critical value of the black hole mass ratio for the violation grows larger.
By comparing the critical value of the black hole mass ratio in the case of $\Lambda  = 0.01$ and $\Lambda  = 0.05$, we find that the increase of the cosmological constant will cause the critical value for the violation to increase.
Put it another way, both the cosmological constant and the loop quantum correction can moderate the violation of the SCC.
However, the violation of SCC still persists as long as the black hole mass ratio is abysmally close to 1.
By comparing the quantum scalar field in RNdS spacetime~\cite{Hollands:2019whz}, we find both the quantum effects in black hole spacetime and the quantum effects in the scalar field can play a key role in restoring the SCC.
However, the quantum effects of the scalar field have a more significant impact on SCC compared to that in black hole spacetime.
The quantum configurations in black hole spacetime can only decelerate the violation of SCC to a certain degree but cannot prevent the occurrence of the violation of SCC.


\section{\label{section6}Concluding remarks}
In this paper, we thoroughly investigate the perturbation of the massless neutral scalar field on a novel loop quantum gravity black hole in de Sitter spacetime.
This model is different from classical black holes in asymptotically flat spacetimes due to the presence of the cosmological constant and the quantum correction parameter. 
Therefore, we focus on studying the influence of parameters in the model on the quasinormal modes and late-time behavior of the scalar field.
Moreover, the global structure between the loop quantum gravity black hole and the charged Reissner-Nordstrom black hole is roughly analogous, so we extend the analysis of the validity of SCC for this quantum-modified black hole in de Sitter spacetime.

To accomplish this purpose, we have performed a time-domain analysis of scalar perturbation and extracted the lowest-lying quasinormal modes by the Prony method.
To validate the reliability of the numerical results obtained, we employed two numerical methods, referred to as matrix method and WKB approximation, for additional calculations.
After all approaches were applied, we obtained consistent results that affirm the accuracy of our data.
On the one hand, we found that the loop quantum gravity black hole exhibits dynamic stability against scalar perturbation.
The effect of loop quantum correction can magnify oscillation frequencies but moderate the decay of the scalar field.
This indicates that the loop quantum effects might be constrained through observing quasinormal frequencies with a long decay time in the ringdown signal.
Moreover, we investigated the effects of the black hole mass ratio and the cosmological constant on the quasinormal spectrum for different multipole numbers. 
It has been observed that as the black hole mass ratio increases, both the real and imaginary parts of the quasinormal modes increase in magnitude.
This suggests that the scalar field undergoes more rapid oscillation and dissipation as the quantum-corrected black hole approaches the extreme state.
When the cosmological constant increases, it has an opposite effect on the quasinormal modes of black holes. 
Instead of speeding up the oscillation and decay rate of scalar fields, it actually slows them down.
On the other hand, the time-domain analysis of scalar perturbation reveals the late-time tail follows an exponential form, which is different from that for asymptotically flat spacetime.
The late-time behavior prominently depends on the multipole number and the cosmological constant.
When the multipole number is zero, the perturbation field decays to a non-zero constant.
As the multipole number increases, the attenuation rate of the late-time tail becomes faster.
However, there is a point where the multipole number is large enough that the late-time response does not display a tail anymore. 
Instead, it shows an infinitely prolonged exponential quasinormal ringing.
Furthermore, we observed that the late-time behavior of the scalar field decays more rapidly as the cosmological constant increases. 
Interestingly, we also found that the effects of the black hole mass ratio and the loop quantum correction on the late-time tail are negligible in comparison.
This suggests that it might be challenging to distinguish the presence of loop quantum corrections in black hole spacetime solely through the analysis of late-time ringdown signal.

In view of the intimate relationship with the stability of the Cauchy horizon and the decay rate of the dominant quasinormal modes, we further explored the validity of the SCC for the quantum-modified black hole under scalar perturbation.
As a result, the SCC is always violated as long as the black hole approaches the extremal one.
Moreover, the violation becomes increasingly challenging with the larger cosmological constant.
The critical value of the black hole mass ratio for the violation increases with the loop quantum correction.
Thus it follows that both the cosmological constant and the loop quantum correction are crucial in mitigating such a violation.
In semiclassical analysis, quantum effects in black hole spacetime have a smaller impact on the SCC compared to quantum effects of quantum scalar fields in RNdS spacetime~\cite{Hollands:2019whz}.
The effect of loop quantum correction in black hole spacetime only serves to slow down the violation of SCC but cannot prevent it from occurring.
To have a deep understanding of this issue, it is better to explore what the emergent classical geometry really looks like in loop quantum gravity coupled to the quantum scalar field. 
But this is utterly beyond the scope of this paper and expected to be reported somewhere else. 


\section*{Acknowledgements}
This work is supported by the National Key ${\rm{R\& D}}$ Program of China under Grant No. 2022YFC2204602 and Grant No. 2021YFC2203001, the Natural Science Foundation of China Grant No. 11925503, Grant No. 12075026, Grant No. 12275022 and Grant No. 12361141825.

\bibliography{mybibfile}

\providecommand{\noopsort}[1]{}\providecommand{\singleletter}[1]{#1}%
\begin{thebibliography}{73}%
\makeatletter
\providecommand \@ifxundefined [1]{%
 \@ifx{#1\undefined}
}%
\providecommand \@ifnum [1]{%
 \ifnum #1\expandafter \@firstoftwo
 \else \expandafter \@secondoftwo
 \fi
}%
\providecommand \@ifx [1]{%
 \ifx #1\expandafter \@firstoftwo
 \else \expandafter \@secondoftwo
 \fi
}%
\providecommand \natexlab [1]{#1}%
\providecommand \enquote  [1]{``#1''}%
\providecommand \bibnamefont  [1]{#1}%
\providecommand \bibfnamefont [1]{#1}%
\providecommand \citenamefont [1]{#1}%
\providecommand \href@noop [0]{\@secondoftwo}%
\providecommand \href [0]{\begingroup \@sanitize@url \@href}%
\providecommand \@href[1]{\@@startlink{#1}\@@href}%
\providecommand \@@href[1]{\endgroup#1\@@endlink}%
\providecommand \@sanitize@url [0]{\catcode `\\12\catcode `\$12\catcode
  `\&12\catcode `\#12\catcode `\^12\catcode `\_12\catcode `\%12\relax}%
\providecommand \@@startlink[1]{}%
\providecommand \@@endlink[0]{}%
\providecommand \url  [0]{\begingroup\@sanitize@url \@url }%
\providecommand \@url [1]{\endgroup\@href {#1}{\urlprefix }}%
\providecommand \urlprefix  [0]{URL }%
\providecommand \Eprint [0]{\href }%
\providecommand \doibase [0]{https://doi.org/}%
\providecommand \selectlanguage [0]{\@gobble}%
\providecommand \bibinfo  [0]{\@secondoftwo}%
\providecommand \bibfield  [0]{\@secondoftwo}%
\providecommand \translation [1]{[#1]}%
\providecommand \BibitemOpen [0]{}%
\providecommand \bibitemStop [0]{}%
\providecommand \bibitemNoStop [0]{.\EOS\space}%
\providecommand \EOS [0]{\spacefactor3000\relax}%
\providecommand \BibitemShut  [1]{\csname bibitem#1\endcsname}%
\let\auto@bib@innerbib\@empty
\bibitem [{\citenamefont {Rovelli}\ and\ \citenamefont
  {Smolin}(1995)}]{Rovelli:1994ge}%
  \BibitemOpen
  \bibfield  {author} {\bibinfo {author} {\bibfnamefont {C.}~\bibnamefont
  {Rovelli}}\ and\ \bibinfo {author} {\bibfnamefont {L.}~\bibnamefont
  {Smolin}},\ }\bibfield  {title} {\bibinfo {title} {{Discreteness of area and
  volume in quantum gravity}},\ }\href
  {https://doi.org/10.1016/0550-3213(95)00150-Q} {\bibfield  {journal}
  {\bibinfo  {journal} {Nucl. Phys. B}\ }\textbf {\bibinfo {volume} {442}},\
  \bibinfo {pages} {593} (\bibinfo {year} {1995})},\ \bibinfo {note} {[Erratum:
  Nucl.Phys.B 456, 753--754 (1995)]},\ \Eprint
  {https://arxiv.org/abs/gr-qc/9411005} {arXiv:gr-qc/9411005} \BibitemShut
  {NoStop}%
\bibitem [{\citenamefont {Ashtekar}\ and\ \citenamefont
  {Lewandowski}(1998)}]{Ashtekar:1997fb}%
  \BibitemOpen
  \bibfield  {author} {\bibinfo {author} {\bibfnamefont {A.}~\bibnamefont
  {Ashtekar}}\ and\ \bibinfo {author} {\bibfnamefont {J.}~\bibnamefont
  {Lewandowski}},\ }\bibfield  {title} {\bibinfo {title} {{Quantum theory of
  geometry. 2. Volume operators}},\ }\href
  {https://doi.org/10.4310/ATMP.1997.v1.n2.a8} {\bibfield  {journal} {\bibinfo
  {journal} {Adv. Theor. Math. Phys.}\ }\textbf {\bibinfo {volume} {1}},\
  \bibinfo {pages} {388} (\bibinfo {year} {1998})},\ \Eprint
  {https://arxiv.org/abs/gr-qc/9711031} {arXiv:gr-qc/9711031} \BibitemShut
  {NoStop}%
\bibitem [{\citenamefont {Han}\ \emph {et~al.}(2007)\citenamefont {Han},
  \citenamefont {Ma},\ and\ \citenamefont {Huang}}]{han2007fundamental}%
  \BibitemOpen
  \bibfield  {author} {\bibinfo {author} {\bibfnamefont {M.}~\bibnamefont
  {Han}}, \bibinfo {author} {\bibfnamefont {Y.}~\bibnamefont {Ma}},\ and\
  \bibinfo {author} {\bibfnamefont {W.}~\bibnamefont {Huang}},\ }\bibfield
  {title} {\bibinfo {title} {Fundamental structure of loop quantum gravity},\
  }\href@noop {} {\bibfield  {journal} {\bibinfo  {journal} {International
  Journal of Modern Physics D}\ }\textbf {\bibinfo {volume} {16}},\ \bibinfo
  {pages} {1397} (\bibinfo {year} {2007})}\BibitemShut {NoStop}%
\bibitem [{\citenamefont {Ashtekar}\ \emph {et~al.}(2006)\citenamefont
  {Ashtekar}, \citenamefont {Pawlowski},\ and\ \citenamefont
  {Singh}}]{ashtekar2006quantumnature}%
  \BibitemOpen
  \bibfield  {author} {\bibinfo {author} {\bibfnamefont {A.}~\bibnamefont
  {Ashtekar}}, \bibinfo {author} {\bibfnamefont {T.}~\bibnamefont
  {Pawlowski}},\ and\ \bibinfo {author} {\bibfnamefont {P.}~\bibnamefont
  {Singh}},\ }\bibfield  {title} {\bibinfo {title} {Quantum nature of the big
  bang: Improved dynamics},\ }\href
  {https://doi.org/10.1103/PhysRevD.74.084003} {\bibfield  {journal} {\bibinfo
  {journal} {Phys. Rev. D}\ }\textbf {\bibinfo {volume} {74}},\ \bibinfo
  {pages} {084003} (\bibinfo {year} {2006})}\BibitemShut {NoStop}%
\bibitem [{\citenamefont {Ashtekar}\ and\ \citenamefont
  {Singh}(2011)}]{ashtekar2011loop}%
  \BibitemOpen
  \bibfield  {author} {\bibinfo {author} {\bibfnamefont {A.}~\bibnamefont
  {Ashtekar}}\ and\ \bibinfo {author} {\bibfnamefont {P.}~\bibnamefont
  {Singh}},\ }\bibfield  {title} {\bibinfo {title} {Loop quantum cosmology: a
  status report},\ }\href@noop {} {\bibfield  {journal} {\bibinfo  {journal}
  {Classical and Quantum Gravity}\ }\textbf {\bibinfo {volume} {28}},\ \bibinfo
  {pages} {213001} (\bibinfo {year} {2011})}\BibitemShut {NoStop}%
\bibitem [{\citenamefont {Zhang}\ \emph
  {et~al.}(2022{\natexlab{a}})\citenamefont {Zhang}, \citenamefont {Song},\
  and\ \citenamefont {Han}}]{Zhang:2021qul}%
  \BibitemOpen
  \bibfield  {author} {\bibinfo {author} {\bibfnamefont {C.}~\bibnamefont
  {Zhang}}, \bibinfo {author} {\bibfnamefont {S.}~\bibnamefont {Song}},\ and\
  \bibinfo {author} {\bibfnamefont {M.}~\bibnamefont {Han}},\ }\bibfield
  {title} {\bibinfo {title} {{First-Order Quantum Correction in Coherent State
  Expectation Value of Loop-Quantum-Gravity Hamiltonian}},\ }\href
  {https://doi.org/10.1103/PhysRevD.105.064008} {\bibfield  {journal} {\bibinfo
   {journal} {Phys. Rev. D}\ }\textbf {\bibinfo {volume} {105}},\ \bibinfo
  {pages} {064008} (\bibinfo {year} {2022}{\natexlab{a}})},\ \Eprint
  {https://arxiv.org/abs/2102.03591} {arXiv:2102.03591 [gr-qc]} \BibitemShut
  {NoStop}%
\bibitem [{\citenamefont {Zhang}\ \emph
  {et~al.}(2022{\natexlab{b}})\citenamefont {Zhang}, \citenamefont {Liu},\ and\
  \citenamefont {Han}}]{Zhang:2022vsl}%
  \BibitemOpen
  \bibfield  {author} {\bibinfo {author} {\bibfnamefont {C.}~\bibnamefont
  {Zhang}}, \bibinfo {author} {\bibfnamefont {H.}~\bibnamefont {Liu}},\ and\
  \bibinfo {author} {\bibfnamefont {M.}~\bibnamefont {Han}},\ }\bibfield
  {title} {\bibinfo {title} {{Fermions in Loop Quantum Gravity and Resolution
  of Doubling Problem}},\ }\href@noop {} {\  (\bibinfo {year}
  {2022}{\natexlab{b}})},\ \Eprint {https://arxiv.org/abs/2212.00933}
  {arXiv:2212.00933 [gr-qc]} \BibitemShut {NoStop}%
\bibitem [{\citenamefont {Chiou}(2008{\natexlab{a}})}]{Chiou:2008nm}%
  \BibitemOpen
  \bibfield  {author} {\bibinfo {author} {\bibfnamefont {D.-W.}\ \bibnamefont
  {Chiou}},\ }\bibfield  {title} {\bibinfo {title} {{Phenomenological loop
  quantum geometry of the Schwarzschild black hole}},\ }\href
  {https://doi.org/10.1103/PhysRevD.78.064040} {\bibfield  {journal} {\bibinfo
  {journal} {Phys. Rev. D}\ }\textbf {\bibinfo {volume} {78}},\ \bibinfo
  {pages} {064040} (\bibinfo {year} {2008}{\natexlab{a}})},\ \Eprint
  {https://arxiv.org/abs/0807.0665} {arXiv:0807.0665 [gr-qc]} \BibitemShut
  {NoStop}%
\bibitem [{\citenamefont {Gambini}\ and\ \citenamefont
  {Pullin}(2008)}]{PhysRevLett.101.161301}%
  \BibitemOpen
  \bibfield  {author} {\bibinfo {author} {\bibfnamefont {R.}~\bibnamefont
  {Gambini}}\ and\ \bibinfo {author} {\bibfnamefont {J.}~\bibnamefont
  {Pullin}},\ }\bibfield  {title} {\bibinfo {title} {Black holes in loop
  quantum gravity: The complete space-time},\ }\href
  {https://doi.org/10.1103/PhysRevLett.101.161301} {\bibfield  {journal}
  {\bibinfo  {journal} {Phys. Rev. Lett.}\ }\textbf {\bibinfo {volume} {101}},\
  \bibinfo {pages} {161301} (\bibinfo {year} {2008})}\BibitemShut {NoStop}%
\bibitem [{\citenamefont {Haggard}\ and\ \citenamefont
  {Rovelli}(2015)}]{Haggard:2014rza}%
  \BibitemOpen
  \bibfield  {author} {\bibinfo {author} {\bibfnamefont {H.~M.}\ \bibnamefont
  {Haggard}}\ and\ \bibinfo {author} {\bibfnamefont {C.}~\bibnamefont
  {Rovelli}},\ }\bibfield  {title} {\bibinfo {title} {{Quantum-gravity effects
  outside the horizon spark black to white hole tunneling}},\ }\href
  {https://doi.org/10.1103/PhysRevD.92.104020} {\bibfield  {journal} {\bibinfo
  {journal} {Phys. Rev. D}\ }\textbf {\bibinfo {volume} {92}},\ \bibinfo
  {pages} {104020} (\bibinfo {year} {2015})},\ \Eprint
  {https://arxiv.org/abs/1407.0989} {arXiv:1407.0989 [gr-qc]} \BibitemShut
  {NoStop}%
\bibitem [{\citenamefont {Christodoulou}\ \emph {et~al.}(2016)\citenamefont
  {Christodoulou}, \citenamefont {Rovelli}, \citenamefont {Speziale},\ and\
  \citenamefont {Vilensky}}]{Christodoulou:2016vny}%
  \BibitemOpen
  \bibfield  {author} {\bibinfo {author} {\bibfnamefont {M.}~\bibnamefont
  {Christodoulou}}, \bibinfo {author} {\bibfnamefont {C.}~\bibnamefont
  {Rovelli}}, \bibinfo {author} {\bibfnamefont {S.}~\bibnamefont {Speziale}},\
  and\ \bibinfo {author} {\bibfnamefont {I.}~\bibnamefont {Vilensky}},\
  }\bibfield  {title} {\bibinfo {title} {{Planck star tunneling time: An
  astrophysically relevant observable from background-free quantum gravity}},\
  }\href {https://doi.org/10.1103/PhysRevD.94.084035} {\bibfield  {journal}
  {\bibinfo  {journal} {Phys. Rev. D}\ }\textbf {\bibinfo {volume} {94}},\
  \bibinfo {pages} {084035} (\bibinfo {year} {2016})},\ \Eprint
  {https://arxiv.org/abs/1605.05268} {arXiv:1605.05268 [gr-qc]} \BibitemShut
  {NoStop}%
\bibitem [{\citenamefont {Ashtekar}\ \emph {et~al.}(2018)\citenamefont
  {Ashtekar}, \citenamefont {Olmedo},\ and\ \citenamefont
  {Singh}}]{Ashtekar:2018lag}%
  \BibitemOpen
  \bibfield  {author} {\bibinfo {author} {\bibfnamefont {A.}~\bibnamefont
  {Ashtekar}}, \bibinfo {author} {\bibfnamefont {J.}~\bibnamefont {Olmedo}},\
  and\ \bibinfo {author} {\bibfnamefont {P.}~\bibnamefont {Singh}},\ }\bibfield
   {title} {\bibinfo {title} {{Quantum Transfiguration of Kruskal Black
  Holes}},\ }\href {https://doi.org/10.1103/PhysRevLett.121.241301} {\bibfield
  {journal} {\bibinfo  {journal} {Phys. Rev. Lett.}\ }\textbf {\bibinfo
  {volume} {121}},\ \bibinfo {pages} {241301} (\bibinfo {year} {2018})},\
  \Eprint {https://arxiv.org/abs/1806.00648} {arXiv:1806.00648 [gr-qc]}
  \BibitemShut {NoStop}%
\bibitem [{\citenamefont {Zhang}\ \emph {et~al.}(2020)\citenamefont {Zhang},
  \citenamefont {Ma}, \citenamefont {Song},\ and\ \citenamefont
  {Zhang}}]{Zhang:2020qxw}%
  \BibitemOpen
  \bibfield  {author} {\bibinfo {author} {\bibfnamefont {C.}~\bibnamefont
  {Zhang}}, \bibinfo {author} {\bibfnamefont {Y.}~\bibnamefont {Ma}}, \bibinfo
  {author} {\bibfnamefont {S.}~\bibnamefont {Song}},\ and\ \bibinfo {author}
  {\bibfnamefont {X.}~\bibnamefont {Zhang}},\ }\bibfield  {title} {\bibinfo
  {title} {{Loop quantum Schwarzschild interior and black hole remnant}},\
  }\href {https://doi.org/10.1103/PhysRevD.102.041502} {\bibfield  {journal}
  {\bibinfo  {journal} {Phys. Rev. D}\ }\textbf {\bibinfo {volume} {102}},\
  \bibinfo {pages} {041502} (\bibinfo {year} {2020})},\ \Eprint
  {https://arxiv.org/abs/2006.08313} {arXiv:2006.08313 [gr-qc]} \BibitemShut
  {NoStop}%
\bibitem [{\citenamefont {Zhang}\ \emph
  {et~al.}(2022{\natexlab{c}})\citenamefont {Zhang}, \citenamefont {Ma},
  \citenamefont {Song},\ and\ \citenamefont {Zhang}}]{Zhang:2021wex}%
  \BibitemOpen
  \bibfield  {author} {\bibinfo {author} {\bibfnamefont {C.}~\bibnamefont
  {Zhang}}, \bibinfo {author} {\bibfnamefont {Y.}~\bibnamefont {Ma}}, \bibinfo
  {author} {\bibfnamefont {S.}~\bibnamefont {Song}},\ and\ \bibinfo {author}
  {\bibfnamefont {X.}~\bibnamefont {Zhang}},\ }\bibfield  {title} {\bibinfo
  {title} {{Loop quantum deparametrized Schwarzschild interior and discrete
  black hole mass}},\ }\href {https://doi.org/10.1103/PhysRevD.105.024069}
  {\bibfield  {journal} {\bibinfo  {journal} {Phys. Rev. D}\ }\textbf {\bibinfo
  {volume} {105}},\ \bibinfo {pages} {024069} (\bibinfo {year}
  {2022}{\natexlab{c}})},\ \Eprint {https://arxiv.org/abs/2107.10579}
  {arXiv:2107.10579 [gr-qc]} \BibitemShut {NoStop}%
\bibitem [{\citenamefont {Lewandowski}\ \emph {et~al.}(2023)\citenamefont
  {Lewandowski}, \citenamefont {Ma}, \citenamefont {Yang},\ and\ \citenamefont
  {Zhang}}]{Lewandowski:2022zce}%
  \BibitemOpen
  \bibfield  {author} {\bibinfo {author} {\bibfnamefont {J.}~\bibnamefont
  {Lewandowski}}, \bibinfo {author} {\bibfnamefont {Y.}~\bibnamefont {Ma}},
  \bibinfo {author} {\bibfnamefont {J.}~\bibnamefont {Yang}},\ and\ \bibinfo
  {author} {\bibfnamefont {C.}~\bibnamefont {Zhang}},\ }\bibfield  {title}
  {\bibinfo {title} {{Quantum Oppenheimer-Snyder and Swiss Cheese Models}},\
  }\href {https://doi.org/10.1103/PhysRevLett.130.101501} {\bibfield  {journal}
  {\bibinfo  {journal} {Phys. Rev. Lett.}\ }\textbf {\bibinfo {volume} {130}},\
  \bibinfo {pages} {101501} (\bibinfo {year} {2023})},\ \Eprint
  {https://arxiv.org/abs/2210.02253} {arXiv:2210.02253 [gr-qc]} \BibitemShut
  {NoStop}%
\bibitem [{\citenamefont {Husain}\ \emph {et~al.}(2022)\citenamefont {Husain},
  \citenamefont {Kelly}, \citenamefont {Santacruz},\ and\ \citenamefont
  {Wilson-Ewing}}]{Husain:2022gwp}%
  \BibitemOpen
  \bibfield  {author} {\bibinfo {author} {\bibfnamefont {V.}~\bibnamefont
  {Husain}}, \bibinfo {author} {\bibfnamefont {J.~G.}\ \bibnamefont {Kelly}},
  \bibinfo {author} {\bibfnamefont {R.}~\bibnamefont {Santacruz}},\ and\
  \bibinfo {author} {\bibfnamefont {E.}~\bibnamefont {Wilson-Ewing}},\
  }\bibfield  {title} {\bibinfo {title} {{Fate of quantum black holes}},\
  }\href {https://doi.org/10.1103/PhysRevD.106.024014} {\bibfield  {journal}
  {\bibinfo  {journal} {Phys. Rev. D}\ }\textbf {\bibinfo {volume} {106}},\
  \bibinfo {pages} {024014} (\bibinfo {year} {2022})},\ \Eprint
  {https://arxiv.org/abs/2203.04238} {arXiv:2203.04238 [gr-qc]} \BibitemShut
  {NoStop}%
\bibitem [{\citenamefont {Han}\ and\ \citenamefont {Liu}(2022)}]{Han:2022rsx}%
  \BibitemOpen
  \bibfield  {author} {\bibinfo {author} {\bibfnamefont {M.}~\bibnamefont
  {Han}}\ and\ \bibinfo {author} {\bibfnamefont {H.}~\bibnamefont {Liu}},\
  }\bibfield  {title} {\bibinfo {title} {{Covariant ${\bar{\mu}}$-scheme
  effective dynamics, mimetic gravity, and non-singular black holes:
  Applications to spherical symmetric quantum gravity and CGHS model}},\
  }\href@noop {} {\  (\bibinfo {year} {2022})},\ \Eprint
  {https://arxiv.org/abs/2212.04605} {arXiv:2212.04605 [gr-qc]} \BibitemShut
  {NoStop}%
\bibitem [{\citenamefont {Han}\ \emph {et~al.}(2023)\citenamefont {Han},
  \citenamefont {Rovelli},\ and\ \citenamefont {Soltani}}]{Han:2023wxg}%
  \BibitemOpen
  \bibfield  {author} {\bibinfo {author} {\bibfnamefont {M.}~\bibnamefont
  {Han}}, \bibinfo {author} {\bibfnamefont {C.}~\bibnamefont {Rovelli}},\ and\
  \bibinfo {author} {\bibfnamefont {F.}~\bibnamefont {Soltani}},\ }\bibfield
  {title} {\bibinfo {title} {{Geometry of the black-to-white hole transition
  within a single asymptotic region}},\ }\href
  {https://doi.org/10.1103/PhysRevD.107.064011} {\bibfield  {journal} {\bibinfo
   {journal} {Phys. Rev. D}\ }\textbf {\bibinfo {volume} {107}},\ \bibinfo
  {pages} {064011} (\bibinfo {year} {2023})},\ \Eprint
  {https://arxiv.org/abs/2302.03872} {arXiv:2302.03872 [gr-qc]} \BibitemShut
  {NoStop}%
\bibitem [{\citenamefont {Stachowiak}\ and\ \citenamefont
  {Szydlowski}(2007)}]{Stachowiak:2006uh}%
  \BibitemOpen
  \bibfield  {author} {\bibinfo {author} {\bibfnamefont {T.}~\bibnamefont
  {Stachowiak}}\ and\ \bibinfo {author} {\bibfnamefont {M.}~\bibnamefont
  {Szydlowski}},\ }\bibfield  {title} {\bibinfo {title} {{Exact solutions in
  bouncing cosmology}},\ }\href
  {https://doi.org/10.1016/j.physletb.2007.01.039} {\bibfield  {journal}
  {\bibinfo  {journal} {Phys. Lett. B}\ }\textbf {\bibinfo {volume} {646}},\
  \bibinfo {pages} {209} (\bibinfo {year} {2007})},\ \Eprint
  {https://arxiv.org/abs/gr-qc/0610121} {arXiv:gr-qc/0610121} \BibitemShut
  {NoStop}%
\bibitem [{\citenamefont {Ashtekar}\ and\ \citenamefont
  {Bojowald}(2006)}]{Ashtekar:2005qt}%
  \BibitemOpen
  \bibfield  {author} {\bibinfo {author} {\bibfnamefont {A.}~\bibnamefont
  {Ashtekar}}\ and\ \bibinfo {author} {\bibfnamefont {M.}~\bibnamefont
  {Bojowald}},\ }\bibfield  {title} {\bibinfo {title} {{Quantum geometry and
  the Schwarzschild singularity}},\ }\href
  {https://doi.org/10.1088/0264-9381/23/2/008} {\bibfield  {journal} {\bibinfo
  {journal} {Class. Quant. Grav.}\ }\textbf {\bibinfo {volume} {23}},\ \bibinfo
  {pages} {391} (\bibinfo {year} {2006})},\ \Eprint
  {https://arxiv.org/abs/gr-qc/0509075} {arXiv:gr-qc/0509075} \BibitemShut
  {NoStop}%
\bibitem [{\citenamefont {Modesto}(2006)}]{Modesto:2005zm}%
  \BibitemOpen
  \bibfield  {author} {\bibinfo {author} {\bibfnamefont {L.}~\bibnamefont
  {Modesto}},\ }\bibfield  {title} {\bibinfo {title} {{Loop quantum black
  hole}},\ }\href {https://doi.org/10.1088/0264-9381/23/18/006} {\bibfield
  {journal} {\bibinfo  {journal} {Class. Quant. Grav.}\ }\textbf {\bibinfo
  {volume} {23}},\ \bibinfo {pages} {5587} (\bibinfo {year} {2006})},\ \Eprint
  {https://arxiv.org/abs/gr-qc/0509078} {arXiv:gr-qc/0509078} \BibitemShut
  {NoStop}%
\bibitem [{\citenamefont {Bojowald}\ and\ \citenamefont
  {Brahma}(2018)}]{Bojowald:2016itl}%
  \BibitemOpen
  \bibfield  {author} {\bibinfo {author} {\bibfnamefont {M.}~\bibnamefont
  {Bojowald}}\ and\ \bibinfo {author} {\bibfnamefont {S.}~\bibnamefont
  {Brahma}},\ }\bibfield  {title} {\bibinfo {title} {{Signature change in
  two-dimensional black-hole models of loop quantum gravity}},\ }\href
  {https://doi.org/10.1103/PhysRevD.98.026012} {\bibfield  {journal} {\bibinfo
  {journal} {Phys. Rev. D}\ }\textbf {\bibinfo {volume} {98}},\ \bibinfo
  {pages} {026012} (\bibinfo {year} {2018})},\ \Eprint
  {https://arxiv.org/abs/1610.08850} {arXiv:1610.08850 [gr-qc]} \BibitemShut
  {NoStop}%
\bibitem [{\citenamefont {Chiou}(2008{\natexlab{b}})}]{Chiou:2008eg}%
  \BibitemOpen
  \bibfield  {author} {\bibinfo {author} {\bibfnamefont {D.-W.}\ \bibnamefont
  {Chiou}},\ }\bibfield  {title} {\bibinfo {title} {{Phenomenological dynamics
  of loop quantum cosmology in Kantowski-Sachs spacetime}},\ }\href
  {https://doi.org/10.1103/PhysRevD.78.044019} {\bibfield  {journal} {\bibinfo
  {journal} {Phys. Rev. D}\ }\textbf {\bibinfo {volume} {78}},\ \bibinfo
  {pages} {044019} (\bibinfo {year} {2008}{\natexlab{b}})},\ \Eprint
  {https://arxiv.org/abs/0803.3659} {arXiv:0803.3659 [gr-qc]} \BibitemShut
  {NoStop}%
\bibitem [{\citenamefont {Zhang}\ \emph {et~al.}(2023)\citenamefont {Zhang},
  \citenamefont {Ma},\ and\ \citenamefont {Yang}}]{Zhang:2023okw}%
  \BibitemOpen
  \bibfield  {author} {\bibinfo {author} {\bibfnamefont {C.}~\bibnamefont
  {Zhang}}, \bibinfo {author} {\bibfnamefont {Y.}~\bibnamefont {Ma}},\ and\
  \bibinfo {author} {\bibfnamefont {J.}~\bibnamefont {Yang}},\ }\bibfield
  {title} {\bibinfo {title} {{Black hole image encoding quantum gravity
  information}},\ }\href {https://doi.org/10.1103/PhysRevD.108.104004}
  {\bibfield  {journal} {\bibinfo  {journal} {Phys. Rev. D}\ }\textbf {\bibinfo
  {volume} {108}},\ \bibinfo {pages} {104004} (\bibinfo {year} {2023})},\
  \Eprint {https://arxiv.org/abs/2302.02800} {arXiv:2302.02800 [gr-qc]}
  \BibitemShut {NoStop}%
\bibitem [{\citenamefont {Yang}\ \emph {et~al.}(2023)\citenamefont {Yang},
  \citenamefont {Zhang},\ and\ \citenamefont {Ma}}]{Yang:2022btw}%
  \BibitemOpen
  \bibfield  {author} {\bibinfo {author} {\bibfnamefont {J.}~\bibnamefont
  {Yang}}, \bibinfo {author} {\bibfnamefont {C.}~\bibnamefont {Zhang}},\ and\
  \bibinfo {author} {\bibfnamefont {Y.}~\bibnamefont {Ma}},\ }\bibfield
  {title} {\bibinfo {title} {{Shadow and stability of quantum-corrected black
  holes}},\ }\href {https://doi.org/10.1140/epjc/s10052-023-11800-8} {\bibfield
   {journal} {\bibinfo  {journal} {Eur. Phys. J. C}\ }\textbf {\bibinfo
  {volume} {83}},\ \bibinfo {pages} {619} (\bibinfo {year} {2023})},\ \Eprint
  {https://arxiv.org/abs/2211.04263} {arXiv:2211.04263 [gr-qc]} \BibitemShut
  {NoStop}%
\bibitem [{\citenamefont {Isi}\ \emph {et~al.}(2019)\citenamefont {Isi},
  \citenamefont {Giesler}, \citenamefont {Farr}, \citenamefont {Scheel},\ and\
  \citenamefont {Teukolsky}}]{Isi:2019aib}%
  \BibitemOpen
  \bibfield  {author} {\bibinfo {author} {\bibfnamefont {M.}~\bibnamefont
  {Isi}}, \bibinfo {author} {\bibfnamefont {M.}~\bibnamefont {Giesler}},
  \bibinfo {author} {\bibfnamefont {W.~M.}\ \bibnamefont {Farr}}, \bibinfo
  {author} {\bibfnamefont {M.~A.}\ \bibnamefont {Scheel}},\ and\ \bibinfo
  {author} {\bibfnamefont {S.~A.}\ \bibnamefont {Teukolsky}},\ }\bibfield
  {title} {\bibinfo {title} {{Testing the no-hair theorem with GW150914}},\
  }\href {https://doi.org/10.1103/PhysRevLett.123.111102} {\bibfield  {journal}
  {\bibinfo  {journal} {Phys. Rev. Lett.}\ }\textbf {\bibinfo {volume} {123}},\
  \bibinfo {pages} {111102} (\bibinfo {year} {2019})},\ \Eprint
  {https://arxiv.org/abs/1905.00869} {arXiv:1905.00869 [gr-qc]} \BibitemShut
  {NoStop}%
\bibitem [{\citenamefont {Gossan}\ \emph {et~al.}(2012)\citenamefont {Gossan},
  \citenamefont {Veitch},\ and\ \citenamefont {Sathyaprakash}}]{Gossan:2011ha}%
  \BibitemOpen
  \bibfield  {author} {\bibinfo {author} {\bibfnamefont {S.}~\bibnamefont
  {Gossan}}, \bibinfo {author} {\bibfnamefont {J.}~\bibnamefont {Veitch}},\
  and\ \bibinfo {author} {\bibfnamefont {B.~S.}\ \bibnamefont
  {Sathyaprakash}},\ }\bibfield  {title} {\bibinfo {title} {{Bayesian model
  selection for testing the no-hair theorem with black hole ringdowns}},\
  }\href {https://doi.org/10.1103/PhysRevD.85.124056} {\bibfield  {journal}
  {\bibinfo  {journal} {Phys. Rev. D}\ }\textbf {\bibinfo {volume} {85}},\
  \bibinfo {pages} {124056} (\bibinfo {year} {2012})},\ \Eprint
  {https://arxiv.org/abs/1111.5819} {arXiv:1111.5819 [gr-qc]} \BibitemShut
  {NoStop}%
\bibitem [{\citenamefont {Ota}\ and\ \citenamefont
  {Chirenti}(2020)}]{Ota:2019bzl}%
  \BibitemOpen
  \bibfield  {author} {\bibinfo {author} {\bibfnamefont {I.}~\bibnamefont
  {Ota}}\ and\ \bibinfo {author} {\bibfnamefont {C.}~\bibnamefont {Chirenti}},\
  }\bibfield  {title} {\bibinfo {title} {{Overtones or higher harmonics?
  Prospects for testing the no-hair theorem with gravitational wave
  detections}},\ }\href {https://doi.org/10.1103/PhysRevD.101.104005}
  {\bibfield  {journal} {\bibinfo  {journal} {Phys. Rev. D}\ }\textbf {\bibinfo
  {volume} {101}},\ \bibinfo {pages} {104005} (\bibinfo {year} {2020})},\
  \Eprint {https://arxiv.org/abs/1911.00440} {arXiv:1911.00440 [gr-qc]}
  \BibitemShut {NoStop}%
\bibitem [{\citenamefont {Carson}\ and\ \citenamefont
  {Yagi}(2020)}]{Carson:2020ter}%
  \BibitemOpen
  \bibfield  {author} {\bibinfo {author} {\bibfnamefont {Z.}~\bibnamefont
  {Carson}}\ and\ \bibinfo {author} {\bibfnamefont {K.}~\bibnamefont {Yagi}},\
  }\bibfield  {title} {\bibinfo {title} {{Probing Einstein-dilaton Gauss-Bonnet
  Gravity with the inspiral and ringdown of gravitational waves}},\ }\href
  {https://doi.org/10.1103/PhysRevD.101.104030} {\bibfield  {journal} {\bibinfo
   {journal} {Phys. Rev. D}\ }\textbf {\bibinfo {volume} {101}},\ \bibinfo
  {pages} {104030} (\bibinfo {year} {2020})},\ \Eprint
  {https://arxiv.org/abs/2003.00286} {arXiv:2003.00286 [gr-qc]} \BibitemShut
  {NoStop}%
\bibitem [{\citenamefont {Shao}\ \emph {et~al.}(2023)\citenamefont {Shao},
  \citenamefont {Hu},\ and\ \citenamefont {Shao}}]{Shao:2023yjx}%
  \BibitemOpen
  \bibfield  {author} {\bibinfo {author} {\bibfnamefont {C.-Y.}\ \bibnamefont
  {Shao}}, \bibinfo {author} {\bibfnamefont {Y.}~\bibnamefont {Hu}},\ and\
  \bibinfo {author} {\bibfnamefont {C.-G.}\ \bibnamefont {Shao}},\ }\bibfield
  {title} {\bibinfo {title} {{Parameter estimation for
  Einstein-dilaton-Gauss-Bonnet gravity with ringdown signals*}},\ }\href
  {https://doi.org/10.1088/1674-1137/ace522} {\bibfield  {journal} {\bibinfo
  {journal} {Chin. Phys. C}\ }\textbf {\bibinfo {volume} {47}},\ \bibinfo
  {pages} {105101} (\bibinfo {year} {2023})},\ \Eprint
  {https://arxiv.org/abs/2307.02084} {arXiv:2307.02084 [gr-qc]} \BibitemShut
  {NoStop}%
\bibitem [{\citenamefont {Bao}\ \emph {et~al.}(2019)\citenamefont {Bao},
  \citenamefont {Shi}, \citenamefont {Wang}, \citenamefont {Zhang},
  \citenamefont {Hu}, \citenamefont {Mei},\ and\ \citenamefont
  {Luo}}]{Bao:2019kgt}%
  \BibitemOpen
  \bibfield  {author} {\bibinfo {author} {\bibfnamefont {J.}~\bibnamefont
  {Bao}}, \bibinfo {author} {\bibfnamefont {C.}~\bibnamefont {Shi}}, \bibinfo
  {author} {\bibfnamefont {H.}~\bibnamefont {Wang}}, \bibinfo {author}
  {\bibfnamefont {J.-d.}\ \bibnamefont {Zhang}}, \bibinfo {author}
  {\bibfnamefont {Y.}~\bibnamefont {Hu}}, \bibinfo {author} {\bibfnamefont
  {J.}~\bibnamefont {Mei}},\ and\ \bibinfo {author} {\bibfnamefont
  {J.}~\bibnamefont {Luo}},\ }\bibfield  {title} {\bibinfo {title}
  {{Constraining modified gravity with ringdown signals: an explicit
  example}},\ }\href {https://doi.org/10.1103/PhysRevD.100.084024} {\bibfield
  {journal} {\bibinfo  {journal} {Phys. Rev. D}\ }\textbf {\bibinfo {volume}
  {100}},\ \bibinfo {pages} {084024} (\bibinfo {year} {2019})},\ \Eprint
  {https://arxiv.org/abs/1905.11674} {arXiv:1905.11674 [gr-qc]} \BibitemShut
  {NoStop}%
\bibitem [{\citenamefont {Berti}\ \emph {et~al.}(2015)\citenamefont {Berti}
  \emph {et~al.}}]{Berti:2015itd}%
  \BibitemOpen
  \bibfield  {author} {\bibinfo {author} {\bibfnamefont {E.}~\bibnamefont
  {Berti}} \emph {et~al.},\ }\bibfield  {title} {\bibinfo {title} {{Testing
  General Relativity with Present and Future Astrophysical Observations}},\
  }\href {https://doi.org/10.1088/0264-9381/32/24/243001} {\bibfield  {journal}
  {\bibinfo  {journal} {Class. Quant. Grav.}\ }\textbf {\bibinfo {volume}
  {32}},\ \bibinfo {pages} {243001} (\bibinfo {year} {2015})},\ \Eprint
  {https://arxiv.org/abs/1501.07274} {arXiv:1501.07274 [gr-qc]} \BibitemShut
  {NoStop}%
\bibitem [{\citenamefont {Konoplya}\ and\ \citenamefont
  {Zhidenko}(2011)}]{RevModPhys.83.793}%
  \BibitemOpen
  \bibfield  {author} {\bibinfo {author} {\bibfnamefont {R.~A.}\ \bibnamefont
  {Konoplya}}\ and\ \bibinfo {author} {\bibfnamefont {A.}~\bibnamefont
  {Zhidenko}},\ }\bibfield  {title} {\bibinfo {title} {Quasinormal modes of
  black holes: From astrophysics to string theory},\ }\href
  {https://doi.org/10.1103/RevModPhys.83.793} {\bibfield  {journal} {\bibinfo
  {journal} {Rev. Mod. Phys.}\ }\textbf {\bibinfo {volume} {83}},\ \bibinfo
  {pages} {793} (\bibinfo {year} {2011})}\BibitemShut {NoStop}%
\bibitem [{\citenamefont {Cardoso}\ \emph
  {et~al.}(2018{\natexlab{a}})\citenamefont {Cardoso}, \citenamefont {Costa},
  \citenamefont {Destounis}, \citenamefont {Hintz},\ and\ \citenamefont
  {Jansen}}]{prl_scc_qnm}%
  \BibitemOpen
  \bibfield  {author} {\bibinfo {author} {\bibfnamefont {V.}~\bibnamefont
  {Cardoso}}, \bibinfo {author} {\bibfnamefont {J.~a.~L.}\ \bibnamefont
  {Costa}}, \bibinfo {author} {\bibfnamefont {K.}~\bibnamefont {Destounis}},
  \bibinfo {author} {\bibfnamefont {P.}~\bibnamefont {Hintz}},\ and\ \bibinfo
  {author} {\bibfnamefont {A.}~\bibnamefont {Jansen}},\ }\bibfield  {title}
  {\bibinfo {title} {Quasinormal modes and strong cosmic censorship},\ }\href
  {https://doi.org/10.1103/PhysRevLett.120.031103} {\bibfield  {journal}
  {\bibinfo  {journal} {Phys. Rev. Lett.}\ }\textbf {\bibinfo {volume} {120}},\
  \bibinfo {pages} {031103} (\bibinfo {year} {2018}{\natexlab{a}})}\BibitemShut
  {NoStop}%
\bibitem [{\citenamefont {Ori}(1999)}]{Ori:1999phc}%
  \BibitemOpen
  \bibfield  {author} {\bibinfo {author} {\bibfnamefont {A.}~\bibnamefont
  {Ori}},\ }\bibfield  {title} {\bibinfo {title} {{Oscillatory null singularity
  inside realistic spinning black holes}},\ }\href
  {https://doi.org/10.1103/PhysRevLett.83.5423} {\bibfield  {journal} {\bibinfo
   {journal} {Phys. Rev. Lett.}\ }\textbf {\bibinfo {volume} {83}},\ \bibinfo
  {pages} {5423} (\bibinfo {year} {1999})},\ \Eprint
  {https://arxiv.org/abs/gr-qc/0103012} {arXiv:gr-qc/0103012} \BibitemShut
  {NoStop}%
\bibitem [{\citenamefont {Poisson}\ and\ \citenamefont
  {Israel}(1990)}]{Poisson:1990eh}%
  \BibitemOpen
  \bibfield  {author} {\bibinfo {author} {\bibfnamefont {E.}~\bibnamefont
  {Poisson}}\ and\ \bibinfo {author} {\bibfnamefont {W.}~\bibnamefont
  {Israel}},\ }\bibfield  {title} {\bibinfo {title} {{Internal structure of
  black holes}},\ }\href {https://doi.org/10.1103/PhysRevD.41.1796} {\bibfield
  {journal} {\bibinfo  {journal} {Phys. Rev. D}\ }\textbf {\bibinfo {volume}
  {41}},\ \bibinfo {pages} {1796} (\bibinfo {year} {1990})}\BibitemShut
  {NoStop}%
\bibitem [{\citenamefont {Destounis}\ \emph
  {et~al.}(2020{\natexlab{a}})\citenamefont {Destounis}, \citenamefont
  {Fontana},\ and\ \citenamefont {Mena}}]{Destounis:2020pjk}%
  \BibitemOpen
  \bibfield  {author} {\bibinfo {author} {\bibfnamefont {K.}~\bibnamefont
  {Destounis}}, \bibinfo {author} {\bibfnamefont {R.~D.~B.}\ \bibnamefont
  {Fontana}},\ and\ \bibinfo {author} {\bibfnamefont {F.~C.}\ \bibnamefont
  {Mena}},\ }\bibfield  {title} {\bibinfo {title} {{Accelerating black holes:
  quasinormal modes and late-time tails}},\ }\href
  {https://doi.org/10.1103/PhysRevD.102.044005} {\bibfield  {journal} {\bibinfo
   {journal} {Phys. Rev. D}\ }\textbf {\bibinfo {volume} {102}},\ \bibinfo
  {pages} {044005} (\bibinfo {year} {2020}{\natexlab{a}})},\ \Eprint
  {https://arxiv.org/abs/2005.03028} {arXiv:2005.03028 [gr-qc]} \BibitemShut
  {NoStop}%
\bibitem [{\citenamefont {Destounis}\ \emph
  {et~al.}(2020{\natexlab{b}})\citenamefont {Destounis}, \citenamefont
  {Fontana},\ and\ \citenamefont {Mena}}]{Destounis:2020yav}%
  \BibitemOpen
  \bibfield  {author} {\bibinfo {author} {\bibfnamefont {K.}~\bibnamefont
  {Destounis}}, \bibinfo {author} {\bibfnamefont {R.~D.~B.}\ \bibnamefont
  {Fontana}},\ and\ \bibinfo {author} {\bibfnamefont {F.~C.}\ \bibnamefont
  {Mena}},\ }\bibfield  {title} {\bibinfo {title} {{Stability of the Cauchy
  horizon in accelerating black-hole spacetimes}},\ }\href
  {https://doi.org/10.1103/PhysRevD.102.104037} {\bibfield  {journal} {\bibinfo
   {journal} {Phys. Rev. D}\ }\textbf {\bibinfo {volume} {102}},\ \bibinfo
  {pages} {104037} (\bibinfo {year} {2020}{\natexlab{b}})},\ \Eprint
  {https://arxiv.org/abs/2006.01152} {arXiv:2006.01152 [gr-qc]} \BibitemShut
  {NoStop}%
\bibitem [{\citenamefont {Zhang}\ and\ \citenamefont {Jiang}(2023)}]{Zhang10}%
  \BibitemOpen
  \bibfield  {author} {\bibinfo {author} {\bibfnamefont {M.}~\bibnamefont
  {Zhang}}\ and\ \bibinfo {author} {\bibfnamefont {J.}~\bibnamefont {Jiang}},\
  }\bibfield  {title} {\bibinfo {title} {{Strong Cosmic Censorship in
  accelerating spacetime}},\ }\href {https://doi.org/10.1007/s11433-023-2117-7}
  {\bibfield  {journal} {\bibinfo  {journal} {Sci. China Phys. Mech. Astron.}\
  }\textbf {\bibinfo {volume} {66}},\ \bibinfo {pages} {280412} (\bibinfo
  {year} {2023})},\ \Eprint {https://arxiv.org/abs/2302.04738}
  {arXiv:2302.04738 [gr-qc]} \BibitemShut {NoStop}%
\bibitem [{\citenamefont {Brady}\ \emph {et~al.}(1999)\citenamefont {Brady},
  \citenamefont {Chambers}, \citenamefont {Laarakkers},\ and\ \citenamefont
  {Poisson}}]{prd_sds_tail}%
  \BibitemOpen
  \bibfield  {author} {\bibinfo {author} {\bibfnamefont {P.~R.}\ \bibnamefont
  {Brady}}, \bibinfo {author} {\bibfnamefont {C.~M.}\ \bibnamefont {Chambers}},
  \bibinfo {author} {\bibfnamefont {W.~G.}\ \bibnamefont {Laarakkers}},\ and\
  \bibinfo {author} {\bibfnamefont {E.}~\bibnamefont {Poisson}},\ }\bibfield
  {title} {\bibinfo {title} {{Radiative falloff in Schwarzschild-de Sitter
  space-time}},\ }\href {https://doi.org/10.1103/PhysRevD.60.064003} {\bibfield
   {journal} {\bibinfo  {journal} {Phys. Rev. D}\ }\textbf {\bibinfo {volume}
  {60}},\ \bibinfo {pages} {064003} (\bibinfo {year} {1999})},\ \Eprint
  {https://arxiv.org/abs/gr-qc/9902010} {arXiv:gr-qc/9902010} \BibitemShut
  {NoStop}%
\bibitem [{\citenamefont {Molina}\ \emph {et~al.}(2004)\citenamefont {Molina},
  \citenamefont {Giugno}, \citenamefont {Abdalla},\ and\ \citenamefont
  {Saa}}]{prd_ds_field}%
  \BibitemOpen
  \bibfield  {author} {\bibinfo {author} {\bibfnamefont {C.}~\bibnamefont
  {Molina}}, \bibinfo {author} {\bibfnamefont {D.}~\bibnamefont {Giugno}},
  \bibinfo {author} {\bibfnamefont {E.}~\bibnamefont {Abdalla}},\ and\ \bibinfo
  {author} {\bibfnamefont {A.}~\bibnamefont {Saa}},\ }\bibfield  {title}
  {\bibinfo {title} {{Field propagation in de Sitter black holes}},\ }\href
  {https://doi.org/10.1103/PhysRevD.69.104013} {\bibfield  {journal} {\bibinfo
  {journal} {Phys. Rev. D}\ }\textbf {\bibinfo {volume} {69}},\ \bibinfo
  {pages} {104013} (\bibinfo {year} {2004})},\ \Eprint
  {https://arxiv.org/abs/gr-qc/0309079} {arXiv:gr-qc/0309079} \BibitemShut
  {NoStop}%
\bibitem [{\citenamefont {Mo}\ \emph {et~al.}(2018)\citenamefont {Mo},
  \citenamefont {Tian}, \citenamefont {Wang}, \citenamefont {Zhang},\ and\
  \citenamefont {Zhong}}]{prd-scc-charg-scal}%
  \BibitemOpen
  \bibfield  {author} {\bibinfo {author} {\bibfnamefont {Y.}~\bibnamefont
  {Mo}}, \bibinfo {author} {\bibfnamefont {Y.}~\bibnamefont {Tian}}, \bibinfo
  {author} {\bibfnamefont {B.}~\bibnamefont {Wang}}, \bibinfo {author}
  {\bibfnamefont {H.}~\bibnamefont {Zhang}},\ and\ \bibinfo {author}
  {\bibfnamefont {Z.}~\bibnamefont {Zhong}},\ }\bibfield  {title} {\bibinfo
  {title} {{Strong cosmic censorship for the massless charged scalar field in
  the Reissner-Nordstrom\textendash{}de Sitter spacetime}},\ }\href
  {https://doi.org/10.1103/PhysRevD.98.124025} {\bibfield  {journal} {\bibinfo
  {journal} {Phys. Rev. D}\ }\textbf {\bibinfo {volume} {98}},\ \bibinfo
  {pages} {124025} (\bibinfo {year} {2018})},\ \Eprint
  {https://arxiv.org/abs/1808.03635} {arXiv:1808.03635 [gr-qc]} \BibitemShut
  {NoStop}%
\bibitem [{\citenamefont {Cardoso}\ \emph
  {et~al.}(2018{\natexlab{b}})\citenamefont {Cardoso}, \citenamefont {Costa},
  \citenamefont {Destounis}, \citenamefont {Hintz},\ and\ \citenamefont
  {Jansen}}]{prd-scc-charg-Cardoso}%
  \BibitemOpen
  \bibfield  {author} {\bibinfo {author} {\bibfnamefont {V.}~\bibnamefont
  {Cardoso}}, \bibinfo {author} {\bibfnamefont {J.~L.}\ \bibnamefont {Costa}},
  \bibinfo {author} {\bibfnamefont {K.}~\bibnamefont {Destounis}}, \bibinfo
  {author} {\bibfnamefont {P.}~\bibnamefont {Hintz}},\ and\ \bibinfo {author}
  {\bibfnamefont {A.}~\bibnamefont {Jansen}},\ }\bibfield  {title} {\bibinfo
  {title} {{Strong cosmic censorship in charged black-hole spacetimes: still
  subtle}},\ }\href {https://doi.org/10.1103/PhysRevD.98.104007} {\bibfield
  {journal} {\bibinfo  {journal} {Phys. Rev. D}\ }\textbf {\bibinfo {volume}
  {98}},\ \bibinfo {pages} {104007} (\bibinfo {year} {2018}{\natexlab{b}})},\
  \Eprint {https://arxiv.org/abs/1808.03631} {arXiv:1808.03631 [gr-qc]}
  \BibitemShut {NoStop}%
\bibitem [{\citenamefont {Dias}\ \emph {et~al.}(2019)\citenamefont {Dias},
  \citenamefont {Reall},\ and\ \citenamefont {Santos}}]{cqg-scc-charge-Harvey}%
  \BibitemOpen
  \bibfield  {author} {\bibinfo {author} {\bibfnamefont {O.~J.~C.}\
  \bibnamefont {Dias}}, \bibinfo {author} {\bibfnamefont {H.~S.}\ \bibnamefont
  {Reall}},\ and\ \bibinfo {author} {\bibfnamefont {J.~E.}\ \bibnamefont
  {Santos}},\ }\bibfield  {title} {\bibinfo {title} {{Strong cosmic censorship
  for charged de Sitter black holes with a charged scalar field}},\ }\href
  {https://doi.org/10.1088/1361-6382/aafcf2} {\bibfield  {journal} {\bibinfo
  {journal} {Class. Quant. Grav.}\ }\textbf {\bibinfo {volume} {36}},\ \bibinfo
  {pages} {045005} (\bibinfo {year} {2019})},\ \Eprint
  {https://arxiv.org/abs/1808.04832} {arXiv:1808.04832 [gr-qc]} \BibitemShut
  {NoStop}%
\bibitem [{\citenamefont {Liu}\ \emph {et~al.}(2019{\natexlab{a}})\citenamefont
  {Liu}, \citenamefont {Tang}, \citenamefont {Destounis}, \citenamefont {Wang},
  \citenamefont {Papantonopoulos},\ and\ \citenamefont
  {Zhang}}]{prd_scc_higher_dimension}%
  \BibitemOpen
  \bibfield  {author} {\bibinfo {author} {\bibfnamefont {H.}~\bibnamefont
  {Liu}}, \bibinfo {author} {\bibfnamefont {Z.}~\bibnamefont {Tang}}, \bibinfo
  {author} {\bibfnamefont {K.}~\bibnamefont {Destounis}}, \bibinfo {author}
  {\bibfnamefont {B.}~\bibnamefont {Wang}}, \bibinfo {author} {\bibfnamefont
  {E.}~\bibnamefont {Papantonopoulos}},\ and\ \bibinfo {author} {\bibfnamefont
  {H.}~\bibnamefont {Zhang}},\ }\bibfield  {title} {\bibinfo {title} {{Strong
  Cosmic Censorship in higher-dimensional Reissner-Nordstr\"om-de Sitter
  spacetime}},\ }\href {https://doi.org/10.1007/JHEP03(2019)187} {\bibfield
  {journal} {\bibinfo  {journal} {JHEP}\ }\textbf {\bibinfo {volume} {03}},\
  \bibinfo {pages} {187}},\ \Eprint {https://arxiv.org/abs/1902.01865}
  {arXiv:1902.01865 [gr-qc]} \BibitemShut {NoStop}%
\bibitem [{\citenamefont {Liu}\ \emph {et~al.}(2019{\natexlab{b}})\citenamefont
  {Liu}, \citenamefont {Van~Vooren}, \citenamefont {Zhang},\ and\ \citenamefont
  {Zhong}}]{jhep_scc_dirac_highter_dimension}%
  \BibitemOpen
  \bibfield  {author} {\bibinfo {author} {\bibfnamefont {X.}~\bibnamefont
  {Liu}}, \bibinfo {author} {\bibfnamefont {S.}~\bibnamefont {Van~Vooren}},
  \bibinfo {author} {\bibfnamefont {H.}~\bibnamefont {Zhang}},\ and\ \bibinfo
  {author} {\bibfnamefont {Z.}~\bibnamefont {Zhong}},\ }\bibfield  {title}
  {\bibinfo {title} {{Strong cosmic censorship for the Dirac field in the
  higher dimensional Reissner-Nordstrom--de Sitter black hole}},\ }\href
  {https://doi.org/10.1007/JHEP10(2019)186} {\bibfield  {journal} {\bibinfo
  {journal} {JHEP}\ }\textbf {\bibinfo {volume} {10}},\ \bibinfo {pages}
  {186}},\ \Eprint {https://arxiv.org/abs/1909.07904} {arXiv:1909.07904
  [hep-th]} \BibitemShut {NoStop}%
\bibitem [{\citenamefont {Ge}\ \emph {et~al.}(2019)\citenamefont {Ge},
  \citenamefont {Jiang}, \citenamefont {Wang}, \citenamefont {Zhang},\ and\
  \citenamefont {Zhong}}]{jhep_scc_dirac_rnds}%
  \BibitemOpen
  \bibfield  {author} {\bibinfo {author} {\bibfnamefont {B.}~\bibnamefont
  {Ge}}, \bibinfo {author} {\bibfnamefont {J.}~\bibnamefont {Jiang}}, \bibinfo
  {author} {\bibfnamefont {B.}~\bibnamefont {Wang}}, \bibinfo {author}
  {\bibfnamefont {H.}~\bibnamefont {Zhang}},\ and\ \bibinfo {author}
  {\bibfnamefont {Z.}~\bibnamefont {Zhong}},\ }\bibfield  {title} {\bibinfo
  {title} {{Strong cosmic censorship for the massless Dirac field in the
  Reissner-Nordstrom-de Sitter spacetime}},\ }\href
  {https://doi.org/10.1007/JHEP01(2019)123} {\bibfield  {journal} {\bibinfo
  {journal} {JHEP}\ }\textbf {\bibinfo {volume} {01}},\ \bibinfo {pages}
  {123}},\ \Eprint {https://arxiv.org/abs/1810.12128} {arXiv:1810.12128
  [gr-qc]} \BibitemShut {NoStop}%
\bibitem [{\citenamefont {Destounis}(2019)}]{plb-charged-fermions}%
  \BibitemOpen
  \bibfield  {author} {\bibinfo {author} {\bibfnamefont {K.}~\bibnamefont
  {Destounis}},\ }\bibfield  {title} {\bibinfo {title} {{Charged Fermions and
  Strong Cosmic Censorship}},\ }\href
  {https://doi.org/10.1016/j.physletb.2019.06.015} {\bibfield  {journal}
  {\bibinfo  {journal} {Phys. Lett. B}\ }\textbf {\bibinfo {volume} {795}},\
  \bibinfo {pages} {211} (\bibinfo {year} {2019})},\ \Eprint
  {https://arxiv.org/abs/1811.10629} {arXiv:1811.10629 [gr-qc]} \BibitemShut
  {NoStop}%
\bibitem [{\citenamefont {Dias}\ \emph
  {et~al.}(2018{\natexlab{a}})\citenamefont {Dias}, \citenamefont {Reall},\
  and\ \citenamefont {Santos}}]{jhep_scc_smooth}%
  \BibitemOpen
  \bibfield  {author} {\bibinfo {author} {\bibfnamefont {O.}~\bibnamefont
  {Dias}}, \bibinfo {author} {\bibfnamefont {H.~S.}\ \bibnamefont {Reall}},\
  and\ \bibinfo {author} {\bibfnamefont {J.~E.}\ \bibnamefont {Santos}},\
  }\bibfield  {title} {\bibinfo {title} {Strong cosmic censorship: taking the
  rough with the smooth},\ }\href@noop {} {\bibfield  {journal} {\bibinfo
  {journal} {Journal of High Energy Physics}\ }\textbf {\bibinfo {volume}
  {2018}},\ \bibinfo {pages} {1} (\bibinfo {year}
  {2018}{\natexlab{a}})}\BibitemShut {NoStop}%
\bibitem [{\citenamefont {Dias}\ \emph
  {et~al.}(2018{\natexlab{b}})\citenamefont {Dias}, \citenamefont {Eperon},
  \citenamefont {Reall},\ and\ \citenamefont {Santos}}]{prd-scc-de-sitt}%
  \BibitemOpen
  \bibfield  {author} {\bibinfo {author} {\bibfnamefont {O.~J.~C.}\
  \bibnamefont {Dias}}, \bibinfo {author} {\bibfnamefont {F.~C.}\ \bibnamefont
  {Eperon}}, \bibinfo {author} {\bibfnamefont {H.~S.}\ \bibnamefont {Reall}},\
  and\ \bibinfo {author} {\bibfnamefont {J.~E.}\ \bibnamefont {Santos}},\
  }\bibfield  {title} {\bibinfo {title} {Strong cosmic censorship in de sitter
  space},\ }\href {https://doi.org/10.1103/PhysRevD.97.104060} {\bibfield
  {journal} {\bibinfo  {journal} {Phys. Rev. D}\ }\textbf {\bibinfo {volume}
  {97}},\ \bibinfo {pages} {104060} (\bibinfo {year}
  {2018}{\natexlab{b}})}\BibitemShut {NoStop}%
\bibitem [{\citenamefont {Rahman}\ \emph {et~al.}(2019)\citenamefont {Rahman},
  \citenamefont {Chakraborty}, \citenamefont {SenGupta},\ and\ \citenamefont
  {Sen}}]{jhep-scc-highspacetime}%
  \BibitemOpen
  \bibfield  {author} {\bibinfo {author} {\bibfnamefont {M.}~\bibnamefont
  {Rahman}}, \bibinfo {author} {\bibfnamefont {S.}~\bibnamefont {Chakraborty}},
  \bibinfo {author} {\bibfnamefont {S.}~\bibnamefont {SenGupta}},\ and\
  \bibinfo {author} {\bibfnamefont {A.~A.}\ \bibnamefont {Sen}},\ }\bibfield
  {title} {\bibinfo {title} {{Fate of Strong Cosmic Censorship Conjecture in
  Presence of Higher Spacetime Dimensions}},\ }\href
  {https://doi.org/10.1007/JHEP03(2019)178} {\bibfield  {journal} {\bibinfo
  {journal} {JHEP}\ }\textbf {\bibinfo {volume} {03}},\ \bibinfo {pages}
  {178}},\ \Eprint {https://arxiv.org/abs/1811.08538} {arXiv:1811.08538
  [gr-qc]} \BibitemShut {NoStop}%
\bibitem [{\citenamefont {Rahman}(2020)}]{epjc-scc-dirac}%
  \BibitemOpen
  \bibfield  {author} {\bibinfo {author} {\bibfnamefont {M.}~\bibnamefont
  {Rahman}},\ }\bibfield  {title} {\bibinfo {title} {{On the validity of Strong
  Cosmic Censorship Conjecture in presence of Dirac fields}},\ }\href
  {https://doi.org/10.1140/epjc/s10052-020-7962-2} {\bibfield  {journal}
  {\bibinfo  {journal} {Eur. Phys. J. C}\ }\textbf {\bibinfo {volume} {80}},\
  \bibinfo {pages} {360} (\bibinfo {year} {2020})},\ \Eprint
  {https://arxiv.org/abs/1905.06675} {arXiv:1905.06675 [gr-qc]} \BibitemShut
  {NoStop}%
\bibitem [{\citenamefont {Casals}\ and\ \citenamefont
  {Marinho}(2022)}]{Casals:2020uxa}%
  \BibitemOpen
  \bibfield  {author} {\bibinfo {author} {\bibfnamefont {M.}~\bibnamefont
  {Casals}}\ and\ \bibinfo {author} {\bibfnamefont {C.~I.~S.}\ \bibnamefont
  {Marinho}},\ }\bibfield  {title} {\bibinfo {title} {{Glimpses of violation of
  strong cosmic censorship in rotating black holes}},\ }\href
  {https://doi.org/10.1103/PhysRevD.106.044060} {\bibfield  {journal} {\bibinfo
   {journal} {Phys. Rev. D}\ }\textbf {\bibinfo {volume} {106}},\ \bibinfo
  {pages} {044060} (\bibinfo {year} {2022})},\ \Eprint
  {https://arxiv.org/abs/2006.06483} {arXiv:2006.06483 [gr-qc]} \BibitemShut
  {NoStop}%
\bibitem [{\citenamefont {Hollands}\ \emph {et~al.}(2020)\citenamefont
  {Hollands}, \citenamefont {Wald},\ and\ \citenamefont
  {Zahn}}]{Hollands:2019whz}%
  \BibitemOpen
  \bibfield  {author} {\bibinfo {author} {\bibfnamefont {S.}~\bibnamefont
  {Hollands}}, \bibinfo {author} {\bibfnamefont {R.~M.}\ \bibnamefont {Wald}},\
  and\ \bibinfo {author} {\bibfnamefont {J.}~\bibnamefont {Zahn}},\ }\bibfield
  {title} {\bibinfo {title} {{Quantum instability of the Cauchy horizon in
  Reissner\textendash{}Nordstr\"om\textendash{}deSitter spacetime}},\ }\href
  {https://doi.org/10.1088/1361-6382/ab8052} {\bibfield  {journal} {\bibinfo
  {journal} {Class. Quant. Grav.}\ }\textbf {\bibinfo {volume} {37}},\ \bibinfo
  {pages} {115009} (\bibinfo {year} {2020})},\ \Eprint
  {https://arxiv.org/abs/1912.06047} {arXiv:1912.06047 [gr-qc]} \BibitemShut
  {NoStop}%
\bibitem [{\citenamefont {Ferrari}\ and\ \citenamefont
  {Mashhoon}(1984)}]{prd-qnm-anal-02}%
  \BibitemOpen
  \bibfield  {author} {\bibinfo {author} {\bibfnamefont {V.}~\bibnamefont
  {Ferrari}}\ and\ \bibinfo {author} {\bibfnamefont {B.}~\bibnamefont
  {Mashhoon}},\ }\bibfield  {title} {\bibinfo {title} {{New approach to the
  quasinormal modes of a black hole}},\ }\href
  {https://doi.org/10.1103/PhysRevD.30.295} {\bibfield  {journal} {\bibinfo
  {journal} {Phys. Rev.}\ }\textbf {\bibinfo {volume} {D30}},\ \bibinfo {pages}
  {295} (\bibinfo {year} {1984})}\BibitemShut {NoStop}%
\bibitem [{\citenamefont {Leaver}(1985)}]{prs-qnm-anal-conti-01}%
  \BibitemOpen
  \bibfield  {author} {\bibinfo {author} {\bibfnamefont {E.~W.}\ \bibnamefont
  {Leaver}},\ }\bibfield  {title} {\bibinfo {title} {{An Analytic
  representation for the quasi normal modes of Kerr black holes}},\ }\href
  {https://doi.org/10.1098/rspa.1985.0119} {\bibfield  {journal} {\bibinfo
  {journal} {Proc. Roy. Soc. Lond.}\ }\textbf {\bibinfo {volume} {A402}},\
  \bibinfo {pages} {285} (\bibinfo {year} {1985})}\BibitemShut {NoStop}%
\bibitem [{\citenamefont {Cho}\ \emph {et~al.}(2012)\citenamefont {Cho},
  \citenamefont {Cornell}, \citenamefont {Doukas}, \citenamefont {Huang},\ and\
  \citenamefont {Naylor}}]{Cho:2011sf}%
  \BibitemOpen
  \bibfield  {author} {\bibinfo {author} {\bibfnamefont {H.~T.}\ \bibnamefont
  {Cho}}, \bibinfo {author} {\bibfnamefont {A.~S.}\ \bibnamefont {Cornell}},
  \bibinfo {author} {\bibfnamefont {J.}~\bibnamefont {Doukas}}, \bibinfo
  {author} {\bibfnamefont {T.~R.}\ \bibnamefont {Huang}},\ and\ \bibinfo
  {author} {\bibfnamefont {W.}~\bibnamefont {Naylor}},\ }\bibfield  {title}
  {\bibinfo {title} {{A New Approach to Black Hole Quasinormal Modes: A Review
  of the Asymptotic Iteration Method}},\ }\href
  {https://doi.org/10.1155/2012/281705} {\bibfield  {journal} {\bibinfo
  {journal} {Adv. Math. Phys.}\ }\textbf {\bibinfo {volume} {2012}},\ \bibinfo
  {pages} {281705} (\bibinfo {year} {2012})},\ \Eprint
  {https://arxiv.org/abs/1111.5024} {arXiv:1111.5024 [gr-qc]} \BibitemShut
  {NoStop}%
\bibitem [{\citenamefont {Gundlach}\ \emph {et~al.}(1994)\citenamefont
  {Gundlach}, \citenamefont {Price},\ and\ \citenamefont
  {Pullin}}]{prd-qnm-lateti-Linear-01}%
  \BibitemOpen
  \bibfield  {author} {\bibinfo {author} {\bibfnamefont {C.}~\bibnamefont
  {Gundlach}}, \bibinfo {author} {\bibfnamefont {R.~H.}\ \bibnamefont
  {Price}},\ and\ \bibinfo {author} {\bibfnamefont {J.}~\bibnamefont
  {Pullin}},\ }\bibfield  {title} {\bibinfo {title} {{Late time behavior of
  stellar collapse and explosions: 1. Linearized perturbations}},\ }\href
  {https://doi.org/10.1103/PhysRevD.49.883} {\bibfield  {journal} {\bibinfo
  {journal} {Phys. Rev.}\ }\textbf {\bibinfo {volume} {D49}},\ \bibinfo {pages}
  {883} (\bibinfo {year} {1994})},\ \Eprint
  {https://arxiv.org/abs/arXiv:gr-qc/9307009} {arXiv:arXiv:gr-qc/9307009
  [gr-qc]} \BibitemShut {NoStop}%
\bibitem [{\citenamefont {Berti}\ \emph {et~al.}(2007)\citenamefont {Berti},
  \citenamefont {Cardoso}, \citenamefont {Gonzalez},\ and\ \citenamefont
  {Sperhake}}]{prd_prony}%
  \BibitemOpen
  \bibfield  {author} {\bibinfo {author} {\bibfnamefont {E.}~\bibnamefont
  {Berti}}, \bibinfo {author} {\bibfnamefont {V.}~\bibnamefont {Cardoso}},
  \bibinfo {author} {\bibfnamefont {J.~A.}\ \bibnamefont {Gonzalez}},\ and\
  \bibinfo {author} {\bibfnamefont {U.}~\bibnamefont {Sperhake}},\ }\bibfield
  {title} {\bibinfo {title} {{Mining information from binary black hole
  mergers: A Comparison of estimation methods for complex exponentials in
  noise}},\ }\href {https://doi.org/10.1103/PhysRevD.75.124017} {\bibfield
  {journal} {\bibinfo  {journal} {Phys. Rev. D}\ }\textbf {\bibinfo {volume}
  {75}},\ \bibinfo {pages} {124017} (\bibinfo {year} {2007})},\ \Eprint
  {https://arxiv.org/abs/gr-qc/0701086} {arXiv:gr-qc/0701086} \BibitemShut
  {NoStop}%
\bibitem [{\citenamefont {Lin}\ and\ \citenamefont
  {Qian}(2017)}]{agr-qnm-lq-matrix-02}%
  \BibitemOpen
  \bibfield  {author} {\bibinfo {author} {\bibfnamefont {K.}~\bibnamefont
  {Lin}}\ and\ \bibinfo {author} {\bibfnamefont {W.-L.}\ \bibnamefont {Qian}},\
  }\bibfield  {title} {\bibinfo {title} {{A Matrix Method for Quasinormal
  Modes: Schwarzschild Black Holes in Asymptotically Flat and (Anti-) de Sitter
  Spacetimes}},\ }\href {https://doi.org/10.1088/1361-6382/aa6643} {\bibfield
  {journal} {\bibinfo  {journal} {Class. Quant. Grav.}\ }\textbf {\bibinfo
  {volume} {34}},\ \bibinfo {pages} {095004} (\bibinfo {year} {2017})},\
  \Eprint {https://arxiv.org/abs/arXiv:1610.08135} {arXiv:arXiv:1610.08135
  [gr-qc]} \BibitemShut {NoStop}%
\bibitem [{\citenamefont {Schutz}\ and\ \citenamefont
  {Will}(1985{\natexlab{a}})}]{aj-nm-semianal-wkb-01}%
  \BibitemOpen
  \bibfield  {author} {\bibinfo {author} {\bibfnamefont {B.~F.}\ \bibnamefont
  {Schutz}}\ and\ \bibinfo {author} {\bibfnamefont {C.~M.}\ \bibnamefont
  {Will}},\ }\bibfield  {title} {\bibinfo {title} {{Black hole normal modes: a
  semianalytic approach}},\ }\href {https://doi.org/10.1086/184453} {\bibfield
  {journal} {\bibinfo  {journal} {Astrophys. J.}\ }\textbf {\bibinfo {volume}
  {291}},\ \bibinfo {pages} {L33} (\bibinfo {year}
  {1985}{\natexlab{a}})}\BibitemShut {NoStop}%
\bibitem [{\citenamefont {Kokkotas}\ and\ \citenamefont
  {Schutz}(1988)}]{prd-nm-wkb-02}%
  \BibitemOpen
  \bibfield  {author} {\bibinfo {author} {\bibfnamefont {K.~D.}\ \bibnamefont
  {Kokkotas}}\ and\ \bibinfo {author} {\bibfnamefont {B.~F.}\ \bibnamefont
  {Schutz}},\ }\bibfield  {title} {\bibinfo {title} {Black-hole normal modes: A
  wkb approach. iii. the reissner-nordstr\"om black hole},\ }\href
  {https://doi.org/10.1103/PhysRevD.37.3378} {\bibfield  {journal} {\bibinfo
  {journal} {Phys. Rev.}\ }\textbf {\bibinfo {volume} {D37}},\ \bibinfo {pages}
  {3378} (\bibinfo {year} {1988})}\BibitemShut {NoStop}%
\bibitem [{\citenamefont {Iyer}\ and\ \citenamefont
  {Will}(1987{\natexlab{a}})}]{prd-nm-wkb-03}%
  \BibitemOpen
  \bibfield  {author} {\bibinfo {author} {\bibfnamefont {S.}~\bibnamefont
  {Iyer}}\ and\ \bibinfo {author} {\bibfnamefont {C.~M.}\ \bibnamefont
  {Will}},\ }\bibfield  {title} {\bibinfo {title} {Black-hole normal modes: A
  wkb approach. i. foundations and application of a higher-order wkb analysis
  of potential-barrier scattering},\ }\href
  {https://doi.org/10.1103/PhysRevD.35.3621} {\bibfield  {journal} {\bibinfo
  {journal} {Phys. Rev.}\ }\textbf {\bibinfo {volume} {D35}},\ \bibinfo {pages}
  {3621} (\bibinfo {year} {1987}{\natexlab{a}})}\BibitemShut {NoStop}%
\bibitem [{\citenamefont {Schutz}\ and\ \citenamefont
  {Will}(1985{\natexlab{b}})}]{article}%
  \BibitemOpen
  \bibfield  {author} {\bibinfo {author} {\bibfnamefont {B.}~\bibnamefont
  {Schutz}}\ and\ \bibinfo {author} {\bibfnamefont {C.}~\bibnamefont {Will}},\
  }\bibfield  {title} {\bibinfo {title} {Black hole normal modes: A
  semianalytic approach},\ }\href {https://doi.org/10.1086/184453} {\bibfield
  {journal} {\bibinfo  {journal} {The Astrophysical Journal, v.291, L33-L36
  (1985)}\ }\textbf {\bibinfo {volume} {291}} (\bibinfo {year}
  {1985}{\natexlab{b}})}\BibitemShut {NoStop}%
\bibitem [{\citenamefont {Iyer}\ and\ \citenamefont
  {Will}(1987{\natexlab{b}})}]{PhysRevD.35.3621}%
  \BibitemOpen
  \bibfield  {author} {\bibinfo {author} {\bibfnamefont {S.}~\bibnamefont
  {Iyer}}\ and\ \bibinfo {author} {\bibfnamefont {C.~M.}\ \bibnamefont
  {Will}},\ }\bibfield  {title} {\bibinfo {title} {Black-hole normal modes: A
  wkb approach. i. foundations and application of a higher-order wkb analysis
  of potential-barrier scattering},\ }\href
  {https://doi.org/10.1103/PhysRevD.35.3621} {\bibfield  {journal} {\bibinfo
  {journal} {Phys. Rev. D}\ }\textbf {\bibinfo {volume} {35}},\ \bibinfo
  {pages} {3621} (\bibinfo {year} {1987}{\natexlab{b}})}\BibitemShut {NoStop}%
\bibitem [{\citenamefont {Iyer}(1987)}]{PhysRevD.35.3632}%
  \BibitemOpen
  \bibfield  {author} {\bibinfo {author} {\bibfnamefont {S.}~\bibnamefont
  {Iyer}},\ }\bibfield  {title} {\bibinfo {title} {Black-hole normal modes: A
  wkb approach. ii. schwarzschild black holes},\ }\href
  {https://doi.org/10.1103/PhysRevD.35.3632} {\bibfield  {journal} {\bibinfo
  {journal} {Phys. Rev. D}\ }\textbf {\bibinfo {volume} {35}},\ \bibinfo
  {pages} {3632} (\bibinfo {year} {1987})}\BibitemShut {NoStop}%
\bibitem [{\citenamefont {Konoplya}(2003)}]{PhysRevD.68.024018}%
  \BibitemOpen
  \bibfield  {author} {\bibinfo {author} {\bibfnamefont {R.~A.}\ \bibnamefont
  {Konoplya}},\ }\bibfield  {title} {\bibinfo {title} {Quasinormal behavior of
  the $d$-dimensional schwarzschild black hole and the higher order wkb
  approach},\ }\href {https://doi.org/10.1103/PhysRevD.68.024018} {\bibfield
  {journal} {\bibinfo  {journal} {Phys. Rev. D}\ }\textbf {\bibinfo {volume}
  {68}},\ \bibinfo {pages} {024018} (\bibinfo {year} {2003})}\BibitemShut
  {NoStop}%
\bibitem [{\citenamefont {Leaver}(1986{\natexlab{a}})}]{Leaver:1986gd}%
  \BibitemOpen
  \bibfield  {author} {\bibinfo {author} {\bibfnamefont {E.~W.}\ \bibnamefont
  {Leaver}},\ }\bibfield  {title} {\bibinfo {title} {{Spectral decomposition of
  the perturbation response of the Schwarzschild geometry}},\ }\href
  {https://doi.org/10.1103/PhysRevD.34.384} {\bibfield  {journal} {\bibinfo
  {journal} {Phys. Rev. D}\ }\textbf {\bibinfo {volume} {34}},\ \bibinfo
  {pages} {384} (\bibinfo {year} {1986}{\natexlab{a}})}\BibitemShut {NoStop}%
\bibitem [{\citenamefont {Ching}\ \emph {et~al.}(1995)\citenamefont {Ching},
  \citenamefont {Leung}, \citenamefont {Suen},\ and\ \citenamefont
  {Young}}]{Ching:1995tj}%
  \BibitemOpen
  \bibfield  {author} {\bibinfo {author} {\bibfnamefont {E.~S.~C.}\
  \bibnamefont {Ching}}, \bibinfo {author} {\bibfnamefont {P.~T.}\ \bibnamefont
  {Leung}}, \bibinfo {author} {\bibfnamefont {W.~M.}\ \bibnamefont {Suen}},\
  and\ \bibinfo {author} {\bibfnamefont {K.}~\bibnamefont {Young}},\ }\bibfield
   {title} {\bibinfo {title} {{Wave propagation in gravitational systems: Late
  time behavior}},\ }\href {https://doi.org/10.1103/PhysRevD.52.2118}
  {\bibfield  {journal} {\bibinfo  {journal} {Phys. Rev. D}\ }\textbf {\bibinfo
  {volume} {52}},\ \bibinfo {pages} {2118} (\bibinfo {year} {1995})},\ \Eprint
  {https://arxiv.org/abs/gr-qc/9507035} {arXiv:gr-qc/9507035} \BibitemShut
  {NoStop}%
\bibitem [{\citenamefont {Leaver}(1986{\natexlab{b}})}]{Leaver:1986vnb}%
  \BibitemOpen
  \bibfield  {author} {\bibinfo {author} {\bibfnamefont {E.~W.}\ \bibnamefont
  {Leaver}},\ }\bibfield  {title} {\bibinfo {title} {{Solutions to a
  generalized spheroidal wave equation: Teukolsky\textquoteright{}s equations
  in general relativity, and the two-center problem in molecular quantum
  mechanics}},\ }\href {https://doi.org/10.1063/1.527130} {\bibfield  {journal}
  {\bibinfo  {journal} {J. Math. Phys.}\ }\textbf {\bibinfo {volume} {27}},\
  \bibinfo {pages} {1238} (\bibinfo {year} {1986}{\natexlab{b}})}\BibitemShut
  {NoStop}%
\bibitem [{\citenamefont {Koyama}\ and\ \citenamefont
  {Tomimatsu}(2001)}]{Koyama:2000hj}%
  \BibitemOpen
  \bibfield  {author} {\bibinfo {author} {\bibfnamefont {H.}~\bibnamefont
  {Koyama}}\ and\ \bibinfo {author} {\bibfnamefont {A.}~\bibnamefont
  {Tomimatsu}},\ }\bibfield  {title} {\bibinfo {title} {{Asymptotic power law
  tails of massive scalar fields in Reissner-Nordstrom background}},\ }\href
  {https://doi.org/10.1103/PhysRevD.63.064032} {\bibfield  {journal} {\bibinfo
  {journal} {Phys. Rev. D}\ }\textbf {\bibinfo {volume} {63}},\ \bibinfo
  {pages} {064032} (\bibinfo {year} {2001})},\ \Eprint
  {https://arxiv.org/abs/gr-qc/0012022} {arXiv:gr-qc/0012022} \BibitemShut
  {NoStop}%
\bibitem [{\citenamefont {Yu}(2002)}]{Yu:2002st}%
  \BibitemOpen
  \bibfield  {author} {\bibinfo {author} {\bibfnamefont {H.-w.}\ \bibnamefont
  {Yu}},\ }\bibfield  {title} {\bibinfo {title} {{Decay of massive scalar hair
  in the background of a black hole with a global monopole}},\ }\href
  {https://doi.org/10.1103/PhysRevD.65.087502} {\bibfield  {journal} {\bibinfo
  {journal} {Phys. Rev. D}\ }\textbf {\bibinfo {volume} {65}},\ \bibinfo
  {pages} {087502} (\bibinfo {year} {2002})},\ \Eprint
  {https://arxiv.org/abs/gr-qc/0201035} {arXiv:gr-qc/0201035} \BibitemShut
  {NoStop}%
\bibitem [{\citenamefont {Poisson}(2002)}]{Poisson:2002jz}%
  \BibitemOpen
  \bibfield  {author} {\bibinfo {author} {\bibfnamefont {E.}~\bibnamefont
  {Poisson}},\ }\bibfield  {title} {\bibinfo {title} {{Radiative falloff of a
  scalar field in a weakly curved space-time without symmetries}},\ }\href
  {https://doi.org/10.1103/PhysRevD.66.044008} {\bibfield  {journal} {\bibinfo
  {journal} {Phys. Rev. D}\ }\textbf {\bibinfo {volume} {66}},\ \bibinfo
  {pages} {044008} (\bibinfo {year} {2002})},\ \Eprint
  {https://arxiv.org/abs/gr-qc/0205018} {arXiv:gr-qc/0205018} \BibitemShut
  {NoStop}%
\end{thebibliography}%

\end{document}